
\input harvmac
%
%
%
%
%
\ifx\answ\bigans
\else
\output={
  \almostshipout{\leftline{\vbox{\pagebody\makefootline}}}
\advancepageno
}
\fi
%
%
%
\def\mayer{\vbox{\sl\centerline{Department of Physics 0319}%
\centerline{University of California, San Diego}
\centerline{9500 Gilman Drive}
\centerline{La Jolla, CA 92093-0319}}}
\def\title#1{\nopagenumbers\hsize=\hsbody%
\centerline{\titlefont #1} \tenpoint \vskip .5in\pageno=0}
%
%
%

%
%
\def\UCSD#1#2{\noindent#1\hfill #2%
\bigskip\supereject\global\hsize=\hsbody%
\footline={\hss\tenrm\folio\hss}}
%
%
\def\abstract#1{\centerline{\bf Abstract}\nobreak\medskip\nobreak\par
#1}
%
%
%
\edef\tfontsize{ scaled\magstep3}
 \tfontsize  \tfontsize
\font\titlermss=cmr5 \tfontsize \font\titlei=cmmi10 \tfontsize
\font\titleis=cmmi7 \tfontsize \font\titleiss=cmmi5 \tfontsize
\font\titlesy=cmsy10 \tfontsize \font\titlesys=cmsy7 \tfontsize
\font\titlesyss=cmsy5 \tfontsize  \tfontsize
\skewchar\titlei='177 \skewchar\titleis='177 \skewchar\titleiss='177
\skewchar\titlesy='60 \skewchar\titlesys='60 \skewchar\titlesyss='60
\scriptscriptfont0=\titlermss
\scriptscriptfont1=\titleiss
\scriptscriptfont2=\titlesyss
%
%
%
%
\def\inv{^{\raise.15ex\hbox{${\scriptscriptstyle -}$}\kern-.05em 1}}
\def\lbar{{\lower.35ex\hbox{$\mathchar'26$}\mkern-10mu\lambda}}

%
%
%
%
\def\dsl{\,\raise.15ex\hbox{/}\mkern-13.5mu D}
\def\delsl{\raise.15ex\hbox{/}\kern-.57em\partial}
\def\Ksl{\hbox{/\kern-.6000em\rm K}}
\def\Asl{\hbox{/\kern-.6500em \rm A}}
\def\Dsl{\hbox{/\kern-.6000em\rm D}} 
\def\Qsl{\hbox{/\kern-.6000em\rm Q}}
\def\gradsl{\hbox{/\kern-.6500em$\nabla$}}
%
%
\def\lspace{\ifx\answ\bigans{}\else\qquad\fi}
\def\lbspace{\ifx\answ\bigans{}\else\hskip-.2in\fi} 
%
%
\def\boxeqn#1{\vcenter{\vbox{\hrule\hbox{\vrule\kern3pt\vbox{\kern3pt
        \hbox{${\displaystyle #1}$}\kern3pt}\kern3pt\vrule}\hrule}}}
%
%
\def\mbox#1#2{\vcenter{\hrule \hbox{\vrule height#2in
\kern#1in \vrule} \hrule}}
%
%
%
%
  \def\CC{{\cal C}} \def\CD{{\cal
D}}
   
   \def\CL{{\cal
L}}
 \def\CN{{\cal N}} \def\CO{{\cal O}}

%
%
%
%
%

%

\def\bar#1{\overline{#1}}
\def\vev#1{\left\langle #1 \right\rangle}

\def\abs#1{\left| #1\right|}

\def\darr#1{\raise1.5ex\hbox{$\leftrightarrow$}\mkern-16.5mu #1}

%
%
\def\half{{\textstyle{1\over2}}}
\def\frac#1#2{{\textstyle{#1\over #2}}} 
%
%
%
%

\def\Tr{\mathop{\rm Tr}}

%
%
%
%

%
%
\def\ltap{\ \raise.3ex\hbox{$<$\kern-.75em\lower1ex\hbox{$\sim$}}\ }
\def\gtap{\ \raise.3ex\hbox{$>$\kern-.75em\lower1ex\hbox{$\sim$}}\ }
\def\gl{\ \raise.5ex\hbox{$>$}\kern-.8em\lower.5ex\hbox{$<$}\ }
\def\roughly#1{\raise.3ex\hbox{$#1$\kern-.75em\lower1ex\hbox{$\sim$}}
}
%
%
\def\ie{\hbox{\it i.e.}}        
\def\eg{\hbox{\it e.g.}}        
\def\etal{\hbox{\it et al.}}

\def\frac#1#2{{\textstyle{#1 \over #2}}}

\def\[{\left[}
\def\]{\right]}
\def\({\left(}
\def\){\right)}

\def\lfm{\medskip\noindent}

\def\CO{{\cal O}}

\def\lta{\ \hbox{\raise.55ex\hbox{$<$}} \!\!\!\!\!
\hbox{\raise-.5ex\hbox{$\sim$}}\ }
\def\gta{\ \hbox{\raise.55ex\hbox{$>$}} \!\!\!\!\!
\hbox{\raise-.5ex\hbox{$\sim$}}\ }
\def\np#1#2#3{{\it Nucl. Phys. \/}{\bf B#1}:#3 (#2)}
\def\pl#1#2#3{{\it Phys. Lett. \/}{\bf {#1}B}:#3 (#2)}
\def\prl#1#2#3{{\it Phys. Rev. Lett. \/}{\bf #1}:#3 (#2)}
\def\physrev#1#2#3{{\it Phys. Rev. \/}{\bf D#1}:#3 (#2)}
\def\ap#1#2#3{{\it Ann. Phys. \/}{\bf #1}:#3 (#2)}
\def\prep#1#2#3{{\it Phys. Rep. \/}{\bf #1}:#3 (#2)}
\def\rmp#1#2#3{{\it Rev. Mod. Phys. \/}{\bf #1}:#3 (#2)}
\def\cmp#1#2#3{{\it Comm. Math. Phys. \/}{\bf #1}:#3 (#2) }
\def\ptp#1#2#3{{\it Prog. Theor. Phys. \/}{#1}:#3 (#2) }

\def\taut{\tau_{_T}}
\def\tauh{\tau_{_H}}
\noblackbox
\vskip 1.in
\centerline{\titlefont{ PROGRESS IN } }
 \medskip
 \title{ ELECTROWEAK BARYOGENESIS}
\bigskip
\centerline{\it To appear in: {\bf Annual Review of Nuclear and
Particle Science, vol. 43} }
\medskip
\centerline{A. G. Cohen\footnote{}{Email:
cohen@andy.bu.edu, dkaplan@ucsd.edu, anelson@ucsd.edu}
\footnote{$^a$}{DOE Outstanding Junior Investigator}}
\centerline{{\sl
Physics Department}} \centerline{{\sl Boston University}}
\centerline{{\sl Boston, MA 02215}}
 \medskip
\centerline{ D. B. Kaplan\footnote{$^b$}{Sloan Fellow}
\footnote{$^c$}{NSF Presidential Young Investigator}  and A. E. Nelson
$^b$\footnote{$^d$}{SSC fellow}}
\bigskip\mayer \bigskip
\centerline{Recent work on generating the excess of matter over antimatter in
the early universe during the electroweak phase transition is reviewed.   }
\bigskip \vfill \UCSD{\hbox{UCSD-PTH-93-02,\ \break
BUHEP-93-4}}{January 1993}
\eject
\centerline{TABLE OF CONTENTS}
\bigskip
\noindent {1.} {INTRODUCTION} \leaderfill{2} \medskip
\noindent {2.} {HIGH TEMPERATURE ELECTROWEAK BARYON VIOLATION} \leaderfill{6}
\medskip  \noindent \quad{2.1.} {Baryon Violation and the Anomaly}
\leaderfill{6}
\medskip  \noindent \quad{2.2.} {Baryon Violation at Zero Temperature}
\leaderfill{7} \medskip  \noindent \quad{2.3.} {Baryon Violation at non-Zero
Temperature} \leaderfill{9} \medskip  \noindent \quad{2.4.} {Rates}
\leaderfill{12}
\medskip  \noindent \quad{2.5.} {Baryon Violation at the Phase Transition}
\leaderfill{15} \medskip  \noindent {3.} {THE WEAK PHASE TRANSITION }
\leaderfill{16} \medskip  \noindent \quad{3.1.} {The Nature of the Transition}
\leaderfill{16} \medskip  \noindent \quad{3.2.}
{Dynamics of the Transition in the
Early Universe} \leaderfill{22} \medskip  \noindent \quad{3.3.} {Propagation
and
Shape of the Bubbles} \leaderfill{23} \medskip  \noindent {4.} {CALCULATING THE
BARYON ASYMMETRY} \leaderfill{26} \medskip  \noindent \quad{4.1.}
{$CP$ violation}
\leaderfill{26} \medskip  \noindent \quad{4.2.} {Timescales} \leaderfill{28}
\medskip
\noindent \quad{4.3.} {The adiabatic ``thick wall'' regime: spontaneous
baryogenesis} \leaderfill{29} \medskip
\noindent \quad{4.4.} {The nonadiabatic ``thin wall'' regime:\hfill\par
the charge transport mechanism} \leaderfill{37} \medskip
\noindent \quad{4.5.} {Conclusions about mechanisms} \leaderfill{41} \medskip
\noindent {5.} {CONSTRAINTS AND EXPERIMENTAL SIGNATURES} \leaderfill{41}
\medskip
\noindent {6.} {FUTURE CALCULATIONS} \leaderfill{43} \medskip
\noindent \quad{ REFERENCES} \leaderfill{44}  \medskip
\noindent \quad{ FIGURE CAPTIONS} \leaderfill{50}

\vfill\eject\newsec{INTRODUCTION}

In the standard hot big bang model, relics from the early universe can give
us much useful information  about  microphysics. For instance the  abundances
of the light elements,   produced when the universe was at   a
temperature of $\sim 1$ MeV,  told us long before the existence of LEP
that    there were at most four, and probably three, species of neutrinos
\ref\kolbturn{Kolb   EW, Turner MS, {\it The Early Universe}, Addison-Wesley
(1990)}.  The most obvious of big bang relics are the baryons, which make
our own existence possible. Furthermore, the universe seems to contain
relatively few   antibaryons.
There is clear evidence that at least the local cluster of galaxies
is made of matter, and there is no plausible mechanism to separate matter
from antimatter on such large scales. The observed abundance of baryons
today implies that when the universe was much hotter than a GeV the ratio
of antibaryons to baryons must have been about one part in $10^8$
\kolbturn. Sakharov  pointed out in 1967 \ref\barcon{ Sakharov  AD, {\it JETP
Lett.}{\bf 5}:24 (1967)} that this cosmological asymmetry between
matter and antimatter could be a calculable result of particle
interactions in the  early universe, teaching us several profound
things about fundamental physics:
\lfm\item{1)} The   number of baryons is not conserved.
\lfm\item{2)}
$C$ (charge conjugation   symmetry) and $CP$ (the product of charge
conjugation symmetry and parity) are   not exact symmetries.
\lfm\item{3)}  The universe must have been out of thermal equilibrium
in order to produce net baryon number, since  the numbers  of baryons
and of antibaryons are equal in thermal equilibrium.  Note that all
known  interactions are in thermal equilibrium when the temperature
of the universe is between  100 GeV  and $10^{12}$ GeV.
\lfm
Sakharov suggested that {\it baryogenesis}  took place  immediately
after the big bang, at a temperature   not far below  the   Planck
scale of
$10^{19}$ GeV, when the universe was expanding so rapidly that
many processes were out of thermal equilibrium.   This scenario was
explicitly
realized with the   advent of grand unified theories (GUTs)
\nref\yosh{Yoshimura  M,
\prl{41}{1978}{281}; E \prl{42}{1979}{7461}}\nref\dimsuss{
Dimopoulos S, Susskind  L, \physrev
{18}{1978}{4500}}\nref\ikkt{Ignatiev AYu, Krasnikov NV,
Kuzmin VA, Tavhelidze AN, \pl{76}{1978}{436}}\nref\ttwz{  Toussaint
D,
Treiman S,   Wilczek  F,  Zee A, \physrev{19}{1979}{1036}}\nref\wein{
Weinberg  S,
\prl{42}{1979}{850}}\nref\weinnan{   Nanopoulos  DV,   Weinberg  S,
\physrev{20}{1979}{2484}}\nref\bsw{   Barr  SM,  Segre  G,  Weldon
HA, \physrev{20}{1979}{2494}}\refs{\yosh-\bsw}, which predict
baryon
number violation, and possibly $CP$ violation, mediated by
ultra-heavy particles~\ref\GG{Georgi H,
Glashow    S, \prl{32}{1974}{438}}.

This ``standard'' GUT baryogenesis scenario is appealingly simple but
does not easily fit into an acceptable  cosmology. One difficulty  is that in
the
standard model of electroweak   interactions baryon number is known
theoretically to be anomalous and not exactly conserved
\ref\thooft{  t'Hooft  G,
\prl{37}{1976}{8}; \physrev{14}{1976}{3432}}. At low temperature this
anomalous baryon violation only proceeds via an exponentially
suppressed tunnelling process, at a rate proportional to $\exp(-4
 \pi/\alpha_{w})\sim 0$. At temperatures above $\sim 100$ GeV,
however,
electroweak baryon violation may proceed rapidly enough to
equilibrate to zero any  baryons produced by grand unification
\nref\lindeone{Linde AD, \pl{70}{1977}{306}
}\refs{\dimsuss,\   \lindeone}. In the
last decade there has been much work indicating that this is the
case~\nref\manton{Manton NS,
\physrev{28}{1983}{2019};   Klinkhammer FR,   Manton NS,
\physrev{30}{1984}{2212}  }\nref\forh{Forgacs P, Horvath Z,
\pl{138}{1984}{397}}\nref\krs{ Kuzmin  VA,    Rubakov  VA,
Shaposhnikov  ME, \pl{155}{1985}{36} }\nref\hightemp{ Arnold P,
McLerran L,
\physrev{36}{1987}{581}; \physrev{37}{1988}{1020} }\nref\klebshaposh{
Khlebnikov SYu, Shaposhnikov ME,
\np{308}{1988}{885}}\nref\amb{Ambjorn J,
Laursen M, Shaposhnikov M, \pl{197}{1989}{49};   Ambjorn J,   Askaard
T,
Porter H,   Shaposhnikov ME, \pl{244}{1990}{479};
\np{353}{1991}{346};
 Ambjorn J,  Farakos K, \pl{294}{1992}{248}}\nref\clmw{Carson L, Li
X, McLerran L, Wang RT, \physrev{42}{1990}{2127}}\nref\dlsfp{    Dine
M, Lechtenfeld O,   Sakita B,   Fischler W, Polchinski J,
\np{342}{1990}{381}
}\refs{\manton-\dlsfp}. However anomalous electroweak processes
conserve $B-L$,
the difference  between baryon and lepton numbers,   and so
a net $B-L$ generated by grand unified or other interactions  will
not   be
erased by electroweak interactions. In the minimal standard model if
$B-L$ is
non zero then the equilibrium baryon number at high temperature is
$B=(12/37)(B-L)$ \nref\aoki{Aoki K-I, \pl{174}{1986}{371}}\nref\kotu{
Kolb  EW, Turner MS, {\it Mod. Phys. Lett.} {\bf A2}:285
(1987)}\nref\krstwo{   Kuzmin  VA,   Rubakov  VA,   Shaposhnikov  ME,
\pl{191}{1987}{171 }   }\nref\harvturn{
Harvey JA, Turner MS, \physrev{42}{1990}{3344}}\nref\nb{Nelson AE,
Barr SM, \pl{246}{1990}{141}}\nref\dolgov{Dolgov AD, \prep{222}{1992}{309
}}\refs{\krs,\ \aoki-\dolgov}\foot{There
has been some confusion    in the literature over this calculation; the
correct fraction is found in \refs{\nb,\ \dolgov}. }.  The possibility
that the observed baryons are a   relic from a net $B-L$ generated at the
GUT scale would be  constrained if   new $B-L$ violating physics (such as
Majorana neutrino masses) were discovered \nref\fuya{Fukugita M, Yanagida
T,   \physrev{42}{1990}{1285}}\nref\cdeo{Campbell
BA, Davidson   S, Ellis F, Olive K,
\pl{256}{1991}{457}}\nref\fglp{Fischler W,
Giudice GF, Leigh   RG, Paban S, \pl{258}{1991}
{45}}\refs{\harvturn,\ \nb,\ \fuya-\fglp}.

 Another problem with GUTs is that they predict a relic abundance of
massive
stable magnetic monopoles much larger than
the observed matter density \nref\zel{ Zel'dovich YaB,
Khlopov MYu, \pl{79}{1978}{239}}\nref\preskill{   Preskill J,
\prl{54}{1979}{1365}}\refs{\zel,\ \preskill}. The simplest GUT models
which
have an experimentally acceptable weak mixing angle  are
supersymmetric \nref\mssm{  Dimopoulos S,  Georgi H,
\np{193}{1981}{150}}\nref\sakai{  Sakai N, {\it Z. Phys.} {\bf
C11}:153 (1981)}\refs{\mssm,\ \sakai}, and in these models the relic
abundance of gravitinos (supersymmetric partners of the graviton) is
incompatible with constraints on the mass density of the universe
\nref\weingravitino{ Weinberg S,
\prl{48}{1982}{1303}}\nref\ellina{ Ellis J,   Linde AD,  Nanopoulos
DV, \pl{118}{1982}{59}}\nref\krauss{Krauss LM,
\np{227}{1983}{556}}\nref\russ{
Khlopov MYu,     Linde AD, \pl{138}{1984}{265}} \nref\elkina{Ellis J,
Kim JE,
Nanopoulos DV,  \pl{145}{1984}{181}}\nref\os{Ovrut BA, Steinhardt PJ,
\pl{147}{1984}{263}}\refs{\weingravitino-\os}. The   most attractive
solution to these problems, which also explains the flatness and
homogeneity
of the observed universe, is  inflation \nref\inflation{  Guth AH,
\physrev{23}{1981}{347}}\nref\linde{Linde AD, {\it Rep. Prog. Phys.}
{\bf 42}:389 (1979) ; \pl{108}{1982}{389}}\nref\albstein{Albrecht A,
Steinhardt   PJ, \prl{48}{1982}{1220}}\refs{\inflation-\albstein},
which greatly  dilutes the abundances of unwanted relics.  However
after an inflationary epoch   any remaining asymmetry between the
numbers of   baryon and antibaryons is  negligible, and the universe
typically reheats to a   temperature which is  less than or of order
$10^{12}$ GeV, well below the scale   of grand unification \kolbturn.
Furthermore gravitino decays may affect nucleosynthesis~\refs{
\russ,\ \elkina} unless the reheat temperature is  below $10^9$ GeV, a
temperature too cool to reinstate the baryon asymmetry through GUT
processes.

There have also been many explanations of the matter-antimatter
asymmetry which do not rely on GUTs \dolgov. Typically these propose
the
existence of new interactions and particles which exist solely in
order to
create   baryons, and which are otherwise unmotivated and
experimentally
untestable.

In 1985 Kuzmin, Rubakov \& Shaposhnikov suggested an elegant and
simple
solution to the  baryogenesis problem \krs. They argued  that
anomalous baryon
number violation  in the standard model of electroweak interactions
is
rapid   at high temperatures and that the weak
phase transition \ref\kirz{Kirzhnitz DA {\it JETP Lett.} {\bf 15}:529
(1972); Kirzhnitz DA, Linde AD, \pl{72}{1972}{471}}, if first order
with
supercooling, provides a natural  way for the universe to get out of
thermal
equilibrium at weak scale  temperatures. Eventually bubbles of the
broken
phase nucleate and expand until they fill the universe; local
departure from
thermal equilibrium takes place in the vicinity of the expanding
bubble
walls.  Since $C$  and $CP$ are   known to be violated by the
electroweak
interactions, it is possible to satisfy   all the Sakharov
baryogenesis
conditions    within the standard model. This suggestion  opened up
the
exciting possibility that we may be able to compute the baryon number
of the universe in terms of  experimentally measurable aspects of the weak
interactions, such as particle masses and $CP$ violating parameters.

The effects of the $CP$  violation in the {\it minimal} standard
model from the
Kobayashi-Maskawa (KM) phase \ref\km{Kobayashi M, Maskawa M,
\ptp{49}{1973}{652}}\ are  much too small to explain  the   observed
baryon
to entropy ratio\foot{Shaposhnikov  has suggested two
conceivable ways to enhance the $CP$ violation in the standard model
at high
temperature \ref\shaposh{ Shaposhnikov ME, {\it  JETP Lett.} {\bf
44}:364
(1986); \np{287}{1987}{757}; \np{299}{1988}{797}; Phys. Lett. 277B
(1992) 324,
Erratum, Phys. Lett. 282B (1992) 483 }; the first
mechanism, dynamical high temperature spontaneous $CP$ violation, is
contradicted by non-perturbative computation \ref\afs{  Ambjorn J,
Farakos K,
Shaposhnikov ME, Niels Bohr Institute preprint NBI-92-20 (1992)}, and
the
second mechanism, reflection of   baryon number from expanding bubble
walls,
according to naive estimates cannot provide a large enough asymmetry.
}.
Explaining this number via baryogenesis in the early universe would
require a
$C$  and $CP$   violating asymmetry larger than  $ 10^{-8}$ in some
out of
equilibrium reaction, while at high temperature $CP$  violation  from
the
KM phase is always multiplied by a function of   small couplings
and mixing angles of order $10^{-20}$
\nref\jarlskog{Jarlskog C, {\it Z.C. Physik} {\bf 29}:491 (1985),
\prl{55}{1985}{1039}}\refs{\shaposh,\ \jarlskog}.  For example, in minimal
SU(5) GUT-scale   baryogenesis with only KM $CP$ violation one  finds a
baryon to photon ratio   $
n_{_B}/n_\gamma\simeq 10^{-20}$ \bsw.
However the feasibility of   baryogenesis during the electroweak
phase transition has been demonstrated in  several simple
{\it non-minimal} versions of the standard model \nref\usone{Cohen AG,
Kaplan DB, Nelson AE,  \pl{245}{1990}{561}; \np{349}{1991}{727}
}\nref\tz{Turok N,  Zadrozny   J,
\prl{65}{1990}{2331}; \np{358}{1991}{471}}\nref\mstv{McLerran L,
Shaposhnikov
M,   Turok  N, Voloshin M, \pl{256}{1991}{451} }\nref\dhss{  Dine M,
Huet P,
Singleton R, Susskind L, \pl{257}{1991}{351} }\nref\usthree{ Cohen
AG,  Kaplan
DB,   Nelson   AE, \pl{263}{1991}{86} }\nref\usfour{Nelson AE,
Kaplan DB,
Cohen AG, \np{373}{1992}{453}}\nref\usfive{Cohen AG,  Kaplan DB,
Nelson AE,
\pl{294}{1992}{57}} \nref\ussix{ Cohen AG, Nelson AE,
\pl{297}{1992}{111}}\refs{\usone-\ussix}. Furthermore the baryon
asymmetry of
the early universe can be used to constrain parameters in  such non-minimal
models;
for instance in the
singlet majoron model \ref\maj{  Chikashige Y,   Mohapatra R,  Peccei
R,
\pl{98}{1981}{265} } an acceptable baryon abundance is
obtainable only if the
mass of the $\tau$ neutrino is greater than 5 MeV \usone.

A crucial test of weak scale baryogenesis comes from the
requirement that any baryons produced during the transition should
survive until the present.  If  anomalous baryon
violating processes occur at a rate which is   faster than the
expansion rate of the universe, then thermal
and chemical equilibrium will be restored  after the transition.
Since the equilibrium
abundances of baryons and antibaryons are equal, any baryon number
which was   created during the transition could be washed out.  In
the
broken phase, the rate of   baryon number
violation is computed  to be proportional to $\exp(- \CO(1)4\pi v/
g T)$, where  $v$ is the value of the order parameter, $g$ is the
weak coupling constant and $T$ is the temperature immediately
after the transition~\refs{\krs,\ \hightemp}. Thus suppression of the
anomalous baryon number violation after the  transition requires a
large jump
in the   Higgs vev $v$ during the transition, \ie
\eqn\washcond{v(T)/T\gtap 1\ \shaposh.}

Requiring that the baryon asymmetry  of the universe be generated
during the
weak transition can give us new information about the $CP$ violating
and Higgs
sectors of the weak interactions,  allowing us to rule out some
models (such as the minimal standard model), and to constrain others.
There is still very little known about the origin
of $CP$ violation or weak symmetry breaking, and any  information we
can
extract from the early universe is welcome.

\newsec{HIGH TEMPERATURE ELECTROWEAK BARYON VIOLATION}

\subsec{Baryon Violation and the Anomaly}
Classically conserved global $U(1)$ charges are well known  to be
potentially
violated in quantum theories with fermions coupled to gauge fields;
this can be
seen through the one loop computation of the divergence of the
corresponding
current which leads to the  anomaly. In the standard model
the
global  baryon and lepton number currents are exactly
conserved at the
classical level. However the anomaly equations give:
\eqn\anom{ \partial_\mu J^\mu_B = \partial_\mu J^\mu_L = N_f ({g^2
\over 32
\pi^2}
W \tilde W - {{g'}^2 \over 32 \pi^2} X\tilde X)}
Here $N_f$ is the number of families, $W_{\mu\nu}$ is the $SU(2)$
field
strength, $X_{\mu\nu}$ the $U_Y(1)$ field strength, and $g$ and $g'$
are the
associated gauge couplings. Note that  the  difference   $B-L$
between the two
charges is strictly conserved. These equations imply that
fluctuations in the
weak field-strength which have a non-zero dot product of the electric
and
magnetic fields will  lead to corresponding fluctuations in baryon
and lepton
number. The right hand side of each of these equations is a total
divergence:
$W\tilde W = \partial_\mu K^\mu$, $X\tilde X=\partial_\mu k^\mu$
where
\eqn\kmu{\eqalign{K^\mu &= \epsilon^{\mu\nu\alpha\beta}(
W^a_{\nu\alpha}
W^a_\beta -{1\over 3} g \epsilon_{abc} W^a_{\nu} W^b_{\alpha}
W^c_\beta)\cr k^\mu &= \epsilon^{\mu\nu\alpha\beta} X_{\nu\alpha}
X_\beta\cr}}
and $W$, $X$ are the gauge potentials associated with the $SU(2)$ and
$U_Y(1)$
groups respectively. Naively this would seem to allow the definition
of a new
baryon current which {\it is} exactly conserved, $J^\mu_B -
N_f g^2/32\pi^2 K^\mu +
N_f {g'}^2 /32\pi^2 k^\mu$. However  $K^\mu$ is not gauge-invariant, and
this
redefinition is not appropriate~\ref\coleman{Coleman S, {\sl Aspects
of
Symmetry}, Cambridge
(1985)}.

If we consider the change in the total baryon number from some
initial
time zero, to some final time $\tau$ (we will always take space to be
a
three-sphere, $S^3$) we can use the anomaly to give the change in the
baryon
number:
\eqn\change{\eqalign{\Delta B &= N_f[N_{CS}(\tau) - N_{CS}(0)] -
N_f[n_{CS}(\tau) - n_{CS}(0)] \cr
N_{CS} &= {g^2\over 32\pi^2}\int d^3x
\epsilon^{ijk}(W_{ij}^a W^a_k - {1\over 3} g \epsilon_{abc} W^a_i
W^b_j
W^c_k)\cr n_{CS} &= {{g'}^2 \over 32 \pi^2} \int d^3x \epsilon^{ijk}
X_{ij} X_k \cr }}
We have written the result in terms of gauge non-invariant objects
$N_{CS}$ and $n_{CS}$, but the {\it difference} at different times of
each of these two objects {\it is} gauge invariant. We are generically
interested in cases where initial and final average values of the gauge
field-strengths are zero, and we wish to know if $\Delta B$ can be non-zero. In
this case   $n_{CS}$ is strictly zero, since it is proportional to the $U_Y(1)$
field   strength, which is zero by assumption;
hence from now on we need only deal with the $SU(2)$ gauge fields
and $N_{CS}$. To compute $\Delta B$ we work in temporal gauge,
$W_0=0$.  Any gauge potential for which the field strength vanishes must then
be
a   time independent gauge transformation of $\vec W = 0$, and so we can
choose our gauge such that $\vec W = N_{CS}=0$ at $t=0$. Then the potential
at $t=\tau$ must be a (time-independent) gauge transformation of
zero:
\eqn\lgt{\vec W (\tau) = h \vec\nabla h^{-1}}
In terms of $h$ we can write the change in baryon number as
\eqn\deltab{
\Delta B= N_f N_{CS}(\tau)
= N_f {g^2\over 32\pi^2} \int d^3x \epsilon^{ijk} \Tr [h\partial_i
h^{-1}h\partial_j  h^{-1}h\partial_k h^{-1}]}

The problem of finding the change in baryon number has been reduced
to
the  classification of the possible functions $h$, which are maps
from $S^3$ into the gauge group $SU(2)$. The parameter $N_{CS}$ in
this
equation is a topological invariant of these maps, known as the
Chern-Simons
number. Thus the baryon number can change if the Chern-Simons number
can be
non-zero. As is well known, the possible values of the Chern-Simons
number for
$SU(2)$ are   the integers, and  so the baryon number can change by
an integral
multiple of the number of families.

\subsec{Baryon Violation at Zero Temperature}
Although we have seen that the baryon number {\it can} change we have
yet to see that it {\it does}, and we must still compute the rate for
baryon violating processes. This requires an understanding of the
dynamics of the fields as they go from the initial to final field
configurations. It is this part of the problem which will require
non-perturbative methods.

Non-perturbative effects in gauge theories began to receive attention
in the
seventies, especially in regard to the famous $U_A(1)$   problem, and
the
problem of baryon violation is not too different from the   problem
of axial
charge violation in QCD~\thooft. Since the $U_Y(1)$ part of the
anomaly does
not play  a significant role, for simplicity we will  consider a pure
$SU(2)$
gauge theory, without the gauged hypercharge. We will also restrict
our
attention to a single family---multiple families will then be easily
dealt
with.

If the standard gauge theory of weak interactions does violate baryon
number by integral units we would expect to find a non-zero amplitude
involving
three quarks combined in a color singlet, plus one  electron (or
electron
neutrino, depending on the electric charge of the three quarks). For
example we
can look at the matrix element $\langle uude \rangle$;
a non-zero value for this object would be a sign of baryon violation.
In the
semi-classical approximation this matrix element is given by a path
integral
dominated by a stationary point, the weak instanton. The matrix
element above
is then evaluated by   replacing the fields by their values at this
stationary
point, and multiplying by the determinant given by the Gaussian
fluctuations
about the instanton. So far, everything is identical to the QCD case.
However,
the weak interaction $SU(2)$ symmetry is spontaneously broken, and
thus  the
instanton is not an exact stationary point; the Euclidean action
coming
from the Higgs potential can be reduced by shrinking the instanton
radius to
zero. However, these configurations are approximate stationary
points, and
should still be the dominant contribution to the functional integral.
The
integral over instanton sizes (including the action from the Higgs
potential)
can  be performed to yield a non-zero matrix element which is
exponentially
small in the weak fine structure constant $\alpha_{w}$ as is
characteristic of
any semi-classical process: $\langle uude \rangle \propto
e^{-2\pi/\alpha_{w}}$.
The smallness of $\alpha_{w}$ implies that such a
matrix element is totally unobservable---baryon violation in the form
of low energy scattering experiments involving small numbers of
particles is
too small to be observed from this source.

The generalization to three families is, as promised, trivial. Since each left
handed doublet
has a zero mode in the instanton background, the only non-vanishing
baryon
violating matrix elements must involve at least one fermion from
each
doublet.  Thus the change in baryon number during any anomalous event
is three
times what it would be for only one family.

\subsec{Baryon Violation at non-Zero Temperature}
The semi-classical picture painted in the previous section
can be thought of as a quantum
tunneling: the instanton represents a barrier penetration, and the
exponential in the coupling is similar to the conventional WKB factor
in quantum mechanics. This picture can be made precise, by
using a canonical Hamiltonian formulation of the gauge theory. The
barrier penetration factor is then an integral along a path in
configuration space of the usual WKB functional, which depends on the
potential energy of the theory.

We can get an idea of how this ``potential'' varies over
configuration space
by  finding its minima. We will first consider only the
potential energy
of the gauge field. One minimum is then easy to find: $W_\mu = 0$ and
no
fermions, which we will
call $\Omega_0$. We can conventionally take the value of this
potential at
$\Omega_0$ to be zero. Other potential minima have gauge fields which
are of the form $W_\mu = h\partial_\mu h^{-1}$. At first sight these
are
not new minima---they are merely gauge transformations of
${\Omega_0}$. However
if we consider a path in configuration space which begins at
$\Omega_0$ and
ends with $W_\mu = h\partial_\mu h^{-1}$, we know from the anomaly
that the
final configuration can  have non-zero baryon and lepton number, both
given  by
the Chern-Simons number of the gauge transformation $h$. Assuming
that $h$ does
have non-zero Chern-Simons number this   configuration is then also a
minimum
of our potential with zero potential energy, but since it has
non-zero baryon
and lepton number is a genuinely different point in configuration
space.
Consequently the potential has an infinite number of minima, which we
can label
as $\Omega_n$, where $n$ is
the change in Chern-Simons number relative to $\Omega_0$ (a gauge
invariant
quantity); note that this is also just the baryon number per family.
All that remains is to find the potential at all other
points of the
infinite-dimensional configuration space---an impossible task.
However
we {\it can} plot a one-dimensional path
in this space which goes through all of these minima,
and we choose the path for which the height of the barrier between
minima is as small as possible. The resulting one-dimensional
potential is
given in \fig\period{The potential energy
of the $SU(2)$ gauge field along the ``sphaleron'' trajectory. The
minima correspond to configurations with zero gauge field strength,
but different baryon number. The lower curve ignores the energy of the
baryons, while the upper curve does not. The zero of energy for the
two curves has been shifted for clarity.}.

The field configuration at the top of the barrier is called the
``sphaleron'' ~\manton, which has potential energy:
\eqn\sphal{E_{\rm sp} = {2 M_W\over \alpha_{w}} B(\lambda/g^2)}
where $B$ is  a constant requiring numerical evaluation; for the
standard model with a single Higgs doublet this parameter  ranges
between
$1.5\le B \le 2.7$ for $\lambda/g^2$ varying between $0$ and
$\infty$ ($\lambda$ is the Higgs self coupling).

Although quantum tunneling through this barrier is irrelevant due
to the small coupling, the finite (although large) barrier height can
have important consequences for baryon violation. So far we have
assumed that
the initial and final configurations in our baryon violating
transitions have
no gauge field, but only an assortment of low energy fermions with
baryon
number differing by  one; this is the case  for the S-matrix element
$\langle
uude\rangle$. But we might also consider processes that have more
complicated
initial and final configurations,  with a total energy {\it greater}
than the
barrier height. In this
case quantum tunneling is not necessary, as we have enough energy
present in the initial configuration to
cross the barrier. This is the
essence of
the finite temperature baryon violation argued by Kuzmin
\etal~\krs---at
temperatures near the sphaleron energy  initial configurations with
energies
above the barrier will be likely
members of the thermal ensemble, this likelihood being given by a
Boltzmann factor, and baryon number changing transitions can occur
without
barrier penetration. At high temperature baryon number changes by
diffusion
across the top of the barrier, with a rate
proportional
to $\exp{(-E_{\rm sp}/T)}$.

We can gain some insight into this problem by imagining a
one-dimensional particle analogue, a pendulum in the gravitational
potential of the   earth. This pendulum  will be  the analog of the
gauge field (or at least one   degree of
freedom of the gauge field). For the baryons, we imagine that each
time the pendulum passes   through an angle of $\pi$ (the point where
the   pendulum points straight up) traveling in the clockwise
direction it moves a   lever which increases a counter by one; when
traveling through $\pi$ in the counterclockwise direction the motion
of the lever decreases the   counter by
one. The configuration space of this  example consists of the   angle
of the pendulum (between $0$ and $2\pi$) and the value of the counter
($B$),   and the classical potential energy of the pendulum is
precisely our  \period.   At zero temperature, barrier penetration
will occur, allowing transitions   among configurations with
different values of the counter; if the barrier is high, these
 processes are suppressed, as are the tunneling events in our gauge
theory. At   finite temperature however, the pendulum will experience
thermal   fluctuations, and at sufficiently high temperature will
frequently cross $\theta=\pi$,   randomly in both directions. Since
each time the pendulum crosses $\pi$ the   counter value
changes,  we see the value of $B$ fluctuate. If we were to start the
system with the counter localized at some point in configuration
space   then $B$ will   diffuse away from this value like
$\sqrt{t}$,   following a typical random walk.

While this picture does show  fluctuations in the counter (and by
analogy local fluctuations in baryon number in the gauge theory
case),  the average value of the counter does not change since fluctuations in
each direction are equally likely; this is not yet the physics of baryon
violation that we are looking for. As emphasized by Kuzmin \etal~\krs, the
average value of the baryon   number can change only if there is some {\it
bias}   which favors fluctuations in one direction over those in the other. The
potential   of \period\ does not show any such bias; fortunately this is the
result   of our leaving out an essential part of the physics. We have so far
plotted   only the potential associated with the gauge field, but
there is of course an energy associated with the baryons  as well. For a
sufficiently dilute system this is just the energy of a free Fermi gas with
fixed   baryon and lepton number, while at high  density it is more
complicated.
In   either case its minimum will be at zero baryon and lepton number, and will
be   symmetric about this value. In our analogy we can  associate a quadratic
potential with the value of the counter,   with a minimum at zero. Thus there
is   some curvature associated with the periodic potential, which we  show in
the upper curve of \period.
This additional effect   provides the bias we are looking
for; if we start the system localized near some value of the baryon
density, it diffuses outwards eventually reaching an equilibrium with zero
average baryon number. The bias of the rate in the direction of increasing
baryon number relative to the decreasing direction is given by  detailed
balance---assuming that this bias is small we may write~\dlsfp:
\eqn\bvrate{{dn_{_B}\over dt} = 3(\Gamma_+ - \Gamma_-) = -3
{\Gamma_a\over
T} \Delta F }
where $\Delta F$ is the free energy  difference between neighboring
minima, and $\Gamma_a$ is the rate per unit volume for fluctuations
between neighboring minima in the absence of bias (\ie\ in the
absence of
fermions). We have assumed that the processes which equilibrate all
degrees of
freedom {\it aside} from the baryon number (and the corresponding
degree of
freedom of the gauge field) are rapid compared to the diffusion time,
and this
is why the free energy appears rather than the internal energy.

\subsec{Rates}
The first serious attempt to compute the anomalous rate $\Gamma_a$
used a technique developed by Langer~\ref\langer{Langer J,
\ap{54}{1969}{258}; {\it
Physica} {\bf 73}:61  (1974)} and Affleck~\ref\affl{Affleck I,
\prl{46}{1981}{388}}  for evaluating the diffusion rate of a system
over a barrier at finite temperature. The procedure effectively
reduces the   theory to one dimension (the degree of freedom we
plotted above), computes the   flux of the system across the barrier
in one direction weighted with a   Boltzmann factor
for each possible initial state, and finally reintroduces the
degrees of freedom transverse to this one mode in a quadratic
approximation with   a thermal occupation. The  idea behind this
technique is to treat the most   important
mode, the one which goes over the barrier most easily, as exactly as
possible, and then use only the small fluctuations around this mode.
The   resulting computation should be valid when the rate of thermal
diffusion across   the barrier is small, and the trajectory over the
sphaleron barrier is   the most important. The result has all the
features that we have described so   far, especially the
characteristic Boltzmann factor associated with the   sphaleron
energy~\refs{\hightemp,\ \clmw}:
\eqn\aandm{\Gamma_a = \gamma (\alpha_{w} T)^{-3} M_W^7
e^{-E_{{\rm sp}}/T}\qquad\qquad {\rm (broken\ phase)}}
Relatively simple arguments have been given for this general form
including the exponential and the $M_W^7$
dependence~\refs{\klebshaposh,\ \dlsfp}.   The
constant $\gamma$  cannot be calculated analytically; it depends   on
the ratio of the Higgs self-coupling to the gauge coupling squared and
has been numerically evaluated in ref.~\clmw. From this form we can
see that the   Boltzmann factor is, naively, a significant
suppression up to temperatures of   hundreds of GeV. Note however
that the energy of the sphaleron configuration is proportional  to
the weak symmetry breaking---if the weak   interactions were
unbroken, we could find a path in configuration space which changes
the baryon number by one unit, {\it with arbitrarily small potential
energy at   each point}. Although the energy in this case is
arbitrarily small, the   integral of the WKB factor along this path
is still bounded below by   $8\pi^2/g^2$, and consequently quantum
tunneling is   suppressed for any value of the   symmetry breaking.
On the other hand, the thermodynamic baryon violation does   not care
about the tunneling factor, but only about the {\it energy} of the
configuration, and consequently in the symmetric phase we can
change   the baryon number with neither suppression from quantum
tunneling, nor   suppression from a small Boltzmann factor. This
means that baryon violation is only    suppressed up to temperatures
where the weak interactions undergo a transition to   the
symmetric phase, which occurs at a temperature around one hundred
GeV. Thus we need the rate of baryon violation not where the Boltzmann
factor is a   large suppression and the calculation of \clmw\  is
applicable, but also in   the symmetric phase where there is no
Boltzmann suppression factor at   all!

The calculation of the baryon violation rate in the symmetric phase
is considerably more difficult than in the broken phase at low
temperatures. The problem is that infrared divergences begin
to invalidate a perturbative expansion when the temperature is
higher than  all mass scales of the theory. These divergences are
actually cut off  by the generation of a ``magnetic screening length''
$\xi_M$   which gives the scale of spatial correlations in the gauge
theory at high   temperature, but
no reliable perturbative scheme involving this screening
exists~\nref\lindetwo{Linde AD, \pl{93}{1980}{327}}\nref\gpy{Gross
DJ, Pisarski
RD, Yaffe L, \rmp{53}{1981}{1}}\refs{\lindetwo,\ \gpy}. A simple
scaling argument allows us to estimate $\xi_M$. At high temperature
spatial
correlations become increasingly classical, and we need only consider
the
classical dynamics of the gauge field. But in this case the
Hamiltonian appears
only in the combination $H/T$. Classically the gauge coupling is an
irrelevant
parameter; by rescaling the fields we can always bring the coupling
out as a
factor $1/\alpha_{w}$ in front of the Hamiltonian. Thus the only
place the
coupling appears is in the combination $\alpha_{w} T$, and therefore
on
dimensional grounds the   screening length is $\xi_M \sim
(\alpha_{w}
T)^{-1}$.  Similarly we can estimate the rate per unit volume of
baryon
violating transitions as
\eqn\dbrate{\Gamma_a = \kappa (\alpha_{w} T)^4\qquad\qquad{\rm
(symmetric\
phase)}}
where $\kappa$ is a dimensionless constant which requires something
beyond a
simple scaling argument to compute.
Ambjorn \etal~\amb\ have attempted to evaluate this rate numerically,
and find
a value  $0.1\ltap \kappa\ltap 1.0$. These computations are
difficult,
suffering from both infrared and ultraviolet divergences; the results
are
noisy, and the diffusive behavior that one expects is not obvious
from the data
(for example, growth of the fluctuations in the Chern-Simons number
as
$\sqrt{t}$ cannot be verified). For these reasons the results must be
taken as
only indicative of the true value.

Recognizing the difficulties of simulating the standard model in
$3+1$
dimensions, Grigoriev, Rubakov \& Shaposhnikov~\ref\grs{Grigoriev D,
Rubakov V, Shaposhnikov M, \pl{216}{1989}{172}\semi
\np{326}{1989}{737}} have   instead
simulated the classical equations of motion for a $1+1$ dimensional
analog of
the standard model, the Abelian Higgs model. This theory has an
anomalous
fermion number, instantons, and sphalerons similar to the $3+1$
dimensional
gauge theory, but  is both easier to simulate, and has better
ultraviolet
behavior than its four dimensional analog. Grigoriev \etal\ used a
Metropolis algorithm to generate an initial field configuration; since the
theory is non-linear and contains an infinite number of degrees of freedom, its
behavior is   hoped to be ergodic, and  this initial configuration should give
``thermal'' results, with a temperature associated with the
statistical ensemble from which it is picked. The classical equations
of motion   were then
used to evolve this configuration forward in time. The result of this
evolution
shows striking confirmation of the thermal diffusion across the
barrier; as
\fig\grig{$N_{CS}$ as a function of time in the Abelian Higgs model.
(From
\grs).}\ shows, the system spends most of   its time near one
(integer) value
of the Chern-Simons number, with only small oscillations around this
value.
However occasionally the system makes a rapid transition to a
neighboring
integer value of $N_{CS}$, and then continues to hover around this
new value.
The rate of these transitions is in good agreement with the $1+1$
dimensional
analog of \aandm, showing the Boltzmann factor associated with
diffusion over a
barrier. In addition, the rate and amplitude of the small
fluctuations around
one of these integral values is measured to be in agreement with
analytic
estimates. Grigoriev \etal\ looked in detail at the field
configurations in the
neighborhood of these rapid transitions and were able to verify that
the field
indeed ``unwinds'' by passing over the barrier between neighboring
minima of
the potential as we have discussed.

\subsec{Baryon Violation at the Phase Transition}

There are several issues which become relevant specifically for
baryon
violation at the electroweak phase transition. If this phase
transition is
first order and  proceeds through the nucleation of bubbles of true
vacuum with
non-zero symmetry breaking amidst a sea of false vacuum, we will then
need to
know the rate of baryon violation not only in the symmetric phase
(given by
\dbrate) or the broken phase when $\exp{(-E_{\rm sp}/T)}$ is small
(given by
\aandm) but also near the interface of the two phases, the bubble
wall. In this
case we need to know how the baryon violation rate makes a transition
between
these two forms. A naive interpolation between these two formul\ae\ as
is done in ref.~\ref\lmt{Liu B,  McLerran L,   Turok N,
\physrev{46}{1992}{2668}}
would indicate that the region of the bubble wall where baryon
violation is significant would be the region where
\eqn\transit{
M_W(T,\phi) \ltap 7 \alpha_{w} T\qquad {\rm(condition\ for\ rapid\ baryon\
violation)}}
Here $M_W(T,\phi)$ is the value of the $W$ boson mass which is
varying through the wall. The rate in this region is still not well known but
is   likely to be similar to the rate outside the bubble. These statements are
not   definitive since the only known explicit computational methods used in
ref.~\clmw\   are not reliable at this point.

An entirely different problem has been raised by Turok
\etal~\nref\turokrev{Turok N, Imperial preprint TP-91-92-33
(1992)}\nref\gts{Grigoriev D, Turok N,  Shaposhnikov M,
\pl{275}{1992}{395}
}\refs{\turokrev,\ \gts}. As the bubble walls created at the weak
phase transition expand, they liberate energy in the conversion of the
false vacuum to the true vacuum. This liberated energy  may move the system
away   from thermal equilibrium in the broken phase behind the bubble wall.
Then,   as is suggested by some simulations in $1+1$ dimensions, none of the
formulas above are directly relevant, since they assume that most of
the degrees of freedom of the system are in equilibrium. Turok \etal\
find that significant Chern-Simons fluctuations persist in this broken phase
region, contrary to the thermal rate predicted from ref.~\aandm. They find that
the rate for these fluctuations actually {\it increases} as the temperature is
lowered. However Dine   has argued qualitatively that this phenomenon
does not occur   in the real $3+1$ dimensional world~\ref\dyale{Dine
M, Proc. 1st Yale-Texas     Workshop ``Baryon Violation at the
Electroweak Scale'', ed. Krauss L, Rey S J,   (World Scientific 1992)}.
At the weak transition the available latent heat is   insufficient to
create a field configuration localized near the top of the sphaleron
barrier---the latent heat in a sphaleron volume is
$\sim(\pi\lambda/g^3)T$,
while the corresponding height of the barrier is $\sim(4\pi/g )T$, and
for a transition in which eq.~\washcond\ is satisfied
$\lambda\ltap g^3$.   This analysis disagrees with ref.~\turokrev\
where the relevant volume   has been taken to be the magnetic
screening volume $(\xi_M)^{3}$   rather than the smaller sphaleron
volume. This interesting possibility will have to be pursued by
numerical investigations in $3+1$ dimensions.

\goodbreak
\newsec{THE WEAK PHASE TRANSITION  }
\subsec{The Nature of the Transition}

The nature of the weak phase transition in the minimal standard model
is fairly well understood in two different limits. The transition is
known to be strongly first order in the Coleman-Weinberg scenario
\nref\cowe{Coleman S,
Weinberg E, \physrev{7}{1973}{1888}}\nref\kilithree{Kirzhnitz DA,
Linde AD,
  \ap{101}{1976}{195}}\nref\guwe{Guth AJ, Weinberg EJ,
\prl{45}{1980}{1131}}\nref\witten{Witten E,
\np{177}{1981}{477}}\nref\heller{Heller UM,
\pl{191}{1987}{109}}\refs{\cowe-\heller}, where the Higgs mass is
tuned so
that at zero temperature the curvature of the effective potential for
the Higgs
vanishes at the origin.  By ``effective potential'', we mean the
free energy density for a spatially homogeneous Higgs field as
computed  in perturbation theory by the methods of Coleman \&
Weinberg~\nref\wewu{Weinberg EJ, Wu A,
\physrev{36}{1987}{2474}}\refs{\cowe,\ \wewu}. For values   of the
Higgs field between two local minima this is not the ``true''
effective potential, which includes spatially inhomogeneous field
configurations and  is always convex~\nref\syma{Symanzik K,
\cmp{16}{1970}{48}}\nref\iim{Iliopoulos
J, Itzykson C, Martin A, \rmp{47}{1975}{165}}\refs{\syma,\ \iim}. The
Coleman
Weinberg effective
potential in principle  is relevant for computing properties of the
phase transition such as the existence of metastable phases and
bubble  nucleation rates~\refs{\langer,\ \wewu}. In the minimal
standard model the Coleman Weinberg scenario requires a Higgs mass of
10 GeV, which is   ruled out;
however in models with several   scalars this scenario is still
viable \nref\ilpa{Iliopoulos J, Papanicolaou N,
\np{111}{1976}{209}}\nref\kuy{Kondo Y,
Umemura I, Yamamoto Y, \pl{263}{1991}{93}}\refs{\ilpa,\ \kuy}. As the
mass of the Higgs increases, the strength of the transition
decreases.   Perturbative
calculations~\nref\doja{Dolan J, Jackiw R,
\physrev{9}{1974}{3320}}\nref\wein{Weinberg S,
\physrev{9}{1974}{3357}}\nref\kilitwo{Kirzhnitz DA, Linde AD, {\it
JETP } {\bf 40}:628 (1974)}\nref\bks{Bochkarev AI, Khlebnikov SYu,
Shaposhnikov ME,
\np{329}{1990}{490}}\nref\anha{Anderson GE,   Hall  LJ,
\physrev{45}{1992}{2685}}\nref\turok{ Turok N,
\prl{68}{1992}{1803}}\nref\carr{Carrington ME,
\physrev{45}{1992}{2933};   University of Minnesota preprint
WIN-92-06,   (1992)}\nref\dlhll{Dine M, Leigh
RL, Huet P, Linde L, Linde D, \pl{283}{1992}{319};
\physrev{46}{1992}{550}}\nref\arno{Arnold P,
\physrev{46}{1992}{2628}}\nref\bbh{Boyd CG,   Brahm DE, Hsu SDH,
University of Chicago preprint  EFI-92-22
(1992)}\nref\ares{Arnold   P, Espinosa E, University of Washington
preprint UW/PT-92-18 (1992) }\nref\badi{Bagnasco JE,  Dine M, UCSC
preprint SCIPP-92-43 (1992)
}\refs{\shaposh,\ \kilithree,\ \heller,\ \wein-\badi}  indicate
a first order transition persists as the Higgs mass increases.
However in the
minimal standard model the lower
bound   on the Higgs mass  is currently   60 GeV,   comparable to the W
mass   \ref\Higgsmass{ Mori T, ``Searches for the neutral Higgs Boson at
LEP'', Talk given   at the XXVI  Rochester International Conference on High
Energy Physics, Dallas,   Texas
August 6-12, (1992)}\nref\gkw{Gleiser M,  Kolb EW, Watkins R,
\np{364}{1991}{411}}\nref\tetr{ Tetradis N, DESY preprint
DESY-91-151, (1991)}\nref\glko{Gleiser M,  Kolb EW,
\prl{69}{1992}{1304}}\nref\glra{
Gleiser M, Ramos RO, Dartmouth preprint DART-HEP-92-08, (1992)}.  For
Higgs masses greater than the gauge boson masses  finite
temperature perturbation theory is unreliable at the critical
temperature~\refs{\linde,\ \carr-\badi,\ \gkw-\glra}.
In the limit  where the Higgs mass is much heavier than the W and
Z   masses, it is a reasonable approximation to neglect the gauge and
Yukawa interactions when studying the transition.  Then the minimal
standard  model is approximately   the linear sigma model for O(4)
symmetry breaking.   The finite temperature phase   transition for the
O(4) model has been studied   numerically
using lattice field theory\nref\lattofour{Jansen K, Seuferling P,
\np{343}{1990}{507}}\nref\ghkpn{ Gavai RV,  Heller UM,  Karsch F,
Neuhaus T,
Plache B \np{ (proc. suppl.)26}{1992}{539};
\pl{294}{1992}{84} }~\refs{
\heller,\ \lattofour,\  \ghkpn}  and
by  imposing an infrared   cutoff which is removed
using the renormalization group    \ref\tewe{Tetradis N,   Wetterich
C, DESY preprint 92-093 (1992)},  and is clearly second order. Thus
for a  heavy Higgs the transition is at most very weakly first order,
as is also  verified by
lattice calculations including the gauge  couplings~\ref\ejk{Evertz
HG, Jersak  J, Kanaya K, \np{285}{1987}{229}}. Computations with N
flavors of   scalars in the   large N limit confirm the result that
the transition is second order or weakly   first
order for a heavy Higgs, and   first order for a
sufficiently light Higgs  \nref\jain{Jain V, MPI preprint MPI-PH/92-72
(1992)}\nref\japa{Jain V, Papadopoulos A, LBL preprint
33067 (1992)}\refs{
\doja,\ \jain,\ \japa}.  It has also been argued using   the
renormalization group (see ref.~\ref\ginsparg{Ginsparg P,
\np{170}{1980}{388}}) that the transition is first order for small
Higgs mass and second order for large Higgs mass
\ref\march{March-Russell J, \pl{296}{1992}{364}}.

The weak phase transition is   under active investigation since
demonstrating
the feasability of electroweak baryogenesis and doing a precise
computation of
the resulting baryon abundance requires an   understanding of the
details of
this transition---the order, the critical
temperature, the bubble nucleation rate,  the velocity and
shape of the
expanding walls, and the Higgs vev after the transition. Most of
these details
depend on the model   of the weak symmetry breaking sector, about
which we
currently have very little experimental information. Requiring a
first order
transition with a sufficiently large Higgs vev after the transition
places
significant constraints on this sector. For
instance, in the
minimal standard model the condition~\washcond\ for avoiding washout
of the
baryon number after the transition places an upper bound on the Higgs
mass
\shaposh. The numerical value of this   bound is still uncertain, but
perturbative calculations place it in the   experimentally ruled out
30--45 GeV
range~\nref\bkus{Bochkarev A, Kuzmin S, Shaposhnikov  M,
\pl{244}{1990}{27}}\nref\dhs{Dine M, Huet P, Singleton B,
\np{375}{1992}{625}
}\refs{\shaposh,\ \bks,\ \dlhll,\ \lmt,\  \bkus,\ \dhs}. In models
with
additional scalars there is a weaker  upper bound on the mass of the
lightest
scalar~\nref\tztwo{Turok N,   Zadrozny J,
\np{369}{1992}{729}}\nref\guid{Giudice GF,
\physrev{45}{1992}{3177}}\nref\myint{Myint S,
\pl{287}{1992}{325}}\nref\laca{
Land D, Carlson ED, \pl{292}{1992}{107}}\nref\suy{Sei N,   Umemura I,
Yamamoto K, Kyoto   preprint NEAP-49 (1992)}\nref\envi{Enqvist K,
Vilja I,
\pl{287}{1992}{119}} \nref\piet{Pietroni M, Padua  preprint,
DFPD-92-TH-36
(1992)}\refs{\kuy,\ \anha,\ \tztwo-\piet}.  For instance Anderson \&
Hall
\anha\ showed that simply adding a gauge singlet scalar $s$  with a $
s^2
\abs{H}^2$ coupling substantially weakens the Higgs mass  bound.
Similarly,
Myint \myint\ showed that in the minimal supersymmetric standard
model, the
coupling of the top squarks to the Higgs doublet weakens the Higgs
mass bound
slightly.

The main theoretical tools   for studying the transition  are finite
temperature perturbation theory and lattice field theory. The lattice
studies
of the finite temperature transition for most of the interesting
cases, such
as for  the minimal standard model  including gauge   boson effects
with a
relatively light Higgs, are still in a  preliminary
stage~\nref\dahe{Damgaard
PH, Heller UM, \pl{171}{1986}{442}; \np{294}{1987}{253};
\np{304}{1988}{63}}\nref\biks{ Bunk  B, Ilgenfritz  EM, Kripfganz  J,
Schiller
A,  \pl{284}{1992}{371}; preprint BI-TP-92-46 (1992)}
\refs{\heller,\ \ejk,\ \dahe,\ \biks}, and do not
contradict  improved  \nref\taka{Takahashi K, {\it Z.
Phys.}{\bf
C26}:601
(1985)}\nref\fendley{Fendley P,
\pl{196}{1987}{175}}\nref\parw{Parwani R,
\physrev{45}{1992}{4695}} one loop perturbation
theory~\refs{  \doja,\ \wein,\ \carr-\bbh,\ \taka-\parw}.
Computing   the effective potential for the order
parameter $\phi$ gives the order of the   transition---if at the
critical
temperature there are two degenerate minima the   transition is first
order. In
\fig\poten{The  effective potential for a theory which has   a first
order
phase transition. The dotted line is the potential at the critical
temperature $T_c$, which has two degenerate minima. The solid line is
the
potential at the low temperature $T_0$, where only the broken
symmetry phase is
stable.   The dashed line shows the potential at an intermediate
temperature
where the broken   phase has lowest
free energy but the symmetric phase is metastable.  The dot-dashed
line is
the potential at high temperatures, where only the symmetric phase is
stable.
  } we plot the effective potential at various temperatures for a
typical
example of a first order phase transition.

The perturbative one loop result for the effective potential  $V$
is~\refs{\cowe,\ \kilithree,\ \doja,\ \wein}
\eqn\effpot{ \eqalign{   V(\phi,T)=&V_{\rm cl}(\phi)+\({1\over 64
\pi^2}\)\sum_i(-1)^{2 s_i}M_i^4(\phi)\ln   M_i^2(\phi)+ ({\rm
polynomial\ in\ }
\phi  ) \cr  &+ \sum_i (-1)^{2s_i}\   T \int   {k^2   d
k\over(2\pi^2) }
           \ln\left[1 -(-1)^{2s_i}
              \exp\( -{ \sqrt{k^2+M_i^2(\phi)} \over T}\)\right]\ ,
\cr}}
where $M_i$ and $s_i$ are   the mass  and spin respectively of the
$i$'th
particle, the sum on $i$ includes all degrees of freedom including
spin,
$V_{\rm cl}$ is the classical potential and the polynomial in $\phi$
is
determined by the choice of renormalization condition. Physically
this result
is   interpreted as the ground state energy plus the free energy of a
gas of
noninteracting particles at finite temperature. We  therefore do not
expect
eq.~\effpot\ to be   accurate at arbitarily high temperatures,  where
the
density of thermal particles is too high to neglect interactions.
Examination
of higher order corrections for   a theory with gauge interactions
and scalar
self coupling $\lambda$ shows that the   naive finite temperature
perturbation
theory at high temperature is an expansion   in powers of $\lambda
T^2/M_\phi^2(\phi)$ and  $g_i^2 T^2/M_i^2(\phi)$, where  $g_i$   is
the
coupling of the $i$'th gauge particle with mass $M_i$ to $\phi$.
Clearly this
expansion is not valid at high temperatures,   or for any values of
$\phi$ for
which there are massless interacting particles. However this one loop
result
can be improved  at high temperatures, by summing up the dominant
high
temperature corrections to all orders. The leading corrections, which
grow as
$T^2$, come from  subgraphs with superficial degree of
divergence 2---\ie\
corrections to the boson propagators~\refs{\doja,\ \wein }.    How
best to
improve perturbation theory is currently a  subject of active
investigation~\nref\eqz{Espinosa JR, Quiros M, Zwirner F,
\pl{191}{1992}{115};
CERN preprint TH.6577/92 (1992)}\nref\bhw{Buchmuller W, Heilbig T,
DESY
preprint 92-117 (1992); Buchmuller W, Heilbig T Walliser D, DESY
preprint
92-151 (1992)}\nref\elmf{Elmfors P, Nordita preprint 92-63-P
(1992)}\nref\ampi{Amelino-Camelia G, Pi S-Y, Boston University
preprint
BUHEP-92-26 (1992)}\refs{\carr-\bbh,\ \taka-\ampi}.  Here we will summarize
the basic
idea of modifying the boson propagators to sum a   class of diagrams
which give
the leading high temperature corrections. For $T$ much larger than
$M_i$ the
integral in eq.~\effpot\ becomes approximately
\eqn\htbos{V_1^{\rm boson} \approx { M_i^2(\phi)   T^2\over24} + {\rm
const+\CO(T)}}
for a boson and
\eqn\htfer{V_1^{\rm fermion} \approx {M_i^2(\phi) T^2\over 48} + {\rm
const}+\CO(1)}
for a fermion, and gives an   effective temperature dependent
contribution to
the scalar mass
\eqn\effmass{\eqalign{{M^2_\phi}_{\rm
eff}= & V_{\rm cl}(\phi)''+\Delta {M^2_\phi}_{\rm eff}\cr
\Delta{M^2_\phi}_{\rm eff}=&\sum_i\[{(3+ (-1)^{2s_i})\over
96}\]{M_i^2}(\phi)''T^2 \ .\cr}}
Including this mass in the scalar propagator is equivalent at one
loop to the
summation of all the leading high temperature contributions of the
``daisy''
graphs such as those in  \fig\daisy{Typical ``daisy'' contribution to
the
effective potential} \refs{\doja,\ \wein }.  In perturbation theory
there is
also a temperature dependent contribution to  the effective mass of
the
longitudinal gauge bosons---the usual Debye mass of order $g T $.
The
transverse components of the gauge bosons are believed to get a non
perturbative contribution to their masses of order $\alpha_{w} T$,
but remain
massless   to all orders in perturbation theory.  A systematic way of
including
the leading $\CO(T^2)$ high
temperature contribution to the $\phi$ propagator
\refs{\ares,\  \parw} is to
add and
subtract from the Lagrangian a term
\eqn\trick{\CL\rightarrow\CL+{\phi^2\over
2}\Delta {M^2_\phi}_{\rm eff}-{\phi^2\over 2}\Delta {M^2_\phi}_{\rm
eff}\ ,}
where the first term is treated as part of the unperturbed
Lagrangian and
included in the scalar propagator while the second term is treated in
perturbation  theory as an interaction which cancels the leading high
temperature pieces of  thermal subloops. A similar but more
complicated method
can be used to sum the leading high temperature contribution to the
gauge boson
propagators (see ref.~\ares).  Improved perturbation theory then
becomes   an
expansion in powers of $\lambda T/{M_\phi(\phi)}_{\rm eff}$ and
$g_i^2 T
/{M_i(\phi)}_{\rm eff}$,  up to logarithms. Because the effective
thermal
masses of the particles grow linearly  with $T$ (for large $T$) the improved
expansion does not
necessarily break down at high temperature. It always breaks down
near a second
order phase transition however, since   ${M_\phi(0)}_{\rm eff}$ goes
to zero as
the critical temperature is approached, and is   also unreliable in
the
vicinity of a very weakly first order transition. For a
sufficiently strongly
first order transition,  it is possible for the expansion parameters
to remain
small in both minima of the effective potential at the critical
temperature.

 For the minimal standard model with a Higgs mass in the (30-80) GeV
range, a
decent approximation to the improved one loop effective   potential
is to
compute eq.~\effpot\ using  the temperature dependent masses for the
bosons
and a high temperature approximation for the integrals, the result is
\dlhll:
\eqn\stpot{ V(\phi,T)\approx{\lambda_T\over 4}\phi^4    -{(2
m_W^3+m_Z^3)T\over 6\pi v_0^3}
\vert \phi\vert^3+{(2 m_W^2+m_Z^2+ 2 m_t^2)(T^2-T_0^2) \over 8
v_0^2}\phi^2
+{\rm const} \ ,}
where
\eqn\tzlt{\eqalign{
T_0^2=&{2 v_0^2\[m_H^2- {3(2m_W^4+m_Z^4-
4m_t^4)/( 8\pi^2 v_0^2)}\]\over (2 m_W^2+m_Z^2+ 2 m_t^2) }\ ,\cr
\lambda_T=&\lambda-{3\over 16 \pi^2 v_o^4}\(  2m_W^4\ln{m_W^2\over
a_BT^2}+m_Z^4\ln{m_Z^2\over a_BT^2}-4m_t^4\ln{m_t^2\over a_FT^2} \)\
,\cr \ln
a_B=&2\ln 4\pi-2   \gamma\approx 3.91\ ,\cr
\ln a_F=&2\ln\pi-2 \gamma\approx 1.14\ .\cr}}
Note the presence of the $\vert \phi\vert ^3$ term, which means that
at the
critical temperature there will be two coexisting minima, hence a
first order
phase  transition \kilithree. For the range of  Higgs masses
considered, this
critical temperature $T_c$ will be just above $T_0$. Is this
calculation
reliable? At the critical temperature, the effective potential has
the form
\eqn\crpt{V(\phi,T_c)\approx{\lambda_{T_c}\over 4}\[\phi-{ (2
m_W^3+m_Z^3)T_c\over
3\pi v_0^3 \lambda_{T_c}} \]^2\phi^2\ .}
Note that the condition \washcond\  requires
\eqn\washcondtwo{\lambda_{T_c}\ltap g^3\ .}
The effective scalar mass in each of the two minima is
\eqn\effmasstwo{{M_\phi}_{\rm eff}\approx\sqrt{1\over 2 \lambda_T}\ {
(2 m_W^3+m_Z^3)T_c\over 3\pi v_0^3}\ . }
In either minimum, additional loops involving the scalar self
coupling will be
suppressed by
\eqn\supp{\sim{\lambda_{T_c} T_c
\over{M_\phi}_{\rm eff}}= {\sqrt{2 \lambda_{T_c}^3}\;3\pi v_0^3\over
(2 m_W^3+m_Z^3)}\sim {m_H^3\over (2 m_W^3+m_Z^3)}\ . }
The  gauge bosons in the broken phase and the longitudinal gauge
bosons in the
symmetric phase are heavier than the scalars and will give rise to
corrections
proportional to powers of $g$. However the transverse
gauge boson masses in the symmetric phase vanish in perturbation
theory; perturbative corrections involving these states become
unsuppressed for $\phi<gT$ and are infrared divergent, making a
perturbative   calculation of the
effective potential at the origin
impossible~\refs{\lindetwo,\ \kilithree,\ \gpy}. If we assume that
these modes receive a nonperturbative mass of   order
$\alpha_w T$ (related to the magnetic screening length), the  uncertainty
in the value of the effective potential at the origin is $\CO(g^6 T^4)$ due
to nonperturbative  effects~\refs{\lindetwo,\ \gpy}.  Nevertheless it
is still possible to establish the order of the transition in
perturbation   theory. If $\lambda_{T_c}$ is small (light Higgs mass)
when compared with $g^2$   then
examination of eq.~\crpt\  shows that the barrier between the two
minima occurs
at relatively large $\phi$ and the height of the   barrier is larger
than the
$g^6 T^4$ uncertainty in the value of the effective   potential in
the symmetric phase. (The calculation of the effective potential also
breaks down near the inflection points, but a lower bound on the barrier height
can be inferred from the  value of the potential in the region between the
minimum and the inflection point where the perturbation expansion parameters
are   small.)
Thus while the nonperturbative effects will induce some uncertainty
into the value of the critical temperature, they will not affect the result
that the transition is first order provided that
\eqn\neccon{   g T_c \ll{ (2 m_W^3+m_Z^3)T_c\over 3\pi v_0^3
\lambda_{T_c}}\ ,
}
which implies  that $ m_H < m_W $.  In the minimal standard model
with a Higgs mass in the   experimentally allowed range eq.~\neccon\ is at best
marginally satisfied, and so perturbative calculations should not  be terribly
accurate. More sophisticated calculations validate this conclusion---for
instance Arnold \& Espinosa \ares\ and Bagnasco \& Dine \badi\ find fairly
large   (40\%)  two loop corrections to the improved one loop result for the
value of the   Higgs vev at the   critical temperature in the minimal standard
model with a 60   GeV Higgs. Clearly better   methods are needed for
quantitative
results,   such as nonperturbative lattice   calculations.

\subsec{Dynamics of the Transition in the Early Universe}

The conventional picture of how a first order cosmological transition
proceeds is that  as the universe expands  it cools in the symmetric phase
until it reaches the critical temperature $T_c$ at which time the broken phase
becomes equally favored. At $T\simeq T_c$  bubbles  of the broken phase
nucleate, but surface tension effects cause these to immediately
shrink, so that the universe remains in an approximately homogeneous state. For
$T$ somewhat below $T_c$, bubbles of broken phase nucleate which are
sufficiently large so that the volume pressure from the lower free energy
inside   the bubble
can overcome the surface tension. These critical bubbles grow until
the universe is completely converted to the broken phase. The time scale
for the formation of a critical bubble  is proportional to $\exp(F_c/T)$,
where $F_c$ is its free energy. At $T=T_c$,  $F_c=\infty$ and so the universe
always supercools,   resulting in a departure from thermal equilibrium
\langer. A detailed study of the bubble growth in the minimal standard model is
given in ref.~\ref\crka{Carrington ME, Kapusta JI, University of Minnesota
preprint TPI-MINN-92/55-T (1992)}.

This picture was recently challenged  by several authors who claimed
that thermal fluctuations in the order parameter could maintain thermal
equilibrium even during a first order transition \refs{\gkw-\glko}, which would
make baryogenesis impossible. For instance  subcritical bubbles of the
broken phase could appear   above the critical temperature, and during the
transition the universe could be an emulsion of both phases  with the fraction
of   broken phase gradually increasing.  Such large thermal fluctations would
also   destroy the validity of  perturbation theory, since perturbative
calculations   assume that the deviations from a homogeneous field
configuration
are small~\nref\gegl{Gelmini G, Gleiser M, UCLA preprint HEP-92-44
(1992)} \refs{\dlhll,\ \glra,\  \gegl}. Fortunately, for those models   in
which eq.~\washcond\ is satisfied such thermal fluctuations are suppressed,
perturbative expansion parameters are less than 1, and the
conventional picture is probably correct~\nref\ande{Anderson G,
\pl{295}{1992}{32} } \refs{\dlhll,\ \glra,\ \gegl,\ \ande}.

Another conventional assumption is that the bubbles are nearly
spherical and that, once they have grown to macrosopic size,  it is a good
approximation to take   the walls to be planar. It has been suggested
\ref\frka{Freese K, Kamionkowski M, \prl{69}{1992}{2743} } that smooth walls
are
unstable towards   developing wrinkles, which would greatly complicate
bubble evolution, but a recent computation  \ref\hkllm{Huet P,
Kajantie K,
Leigh RG, Liu B-H,   McLerran L, SLAC preprint SLAC-PUB-5943 (1992)}
indicates
that for the
weak transition  smooth walls are stable  and that surface tension
will keep
the expanding bubbles  spherical until they collide.

\subsec{Propagation and Shape of the Bubbles}

In the conventional picture of the transition local departure from
thermal
equilibrium, and baryogenesis, primarily occurs in the vicinity of
the expanding bubble walls.  A quantitative
calculation of the
baryon number of the universe  as described in the next section
requires
the temperature at which bubbles nucleate, the wall profile, and
the wall
velocity. If the effective action is known, computing the nucleation
temperature and the  wall profile is a straightforward exercise
\nref\lindethree{Linde AD, \pl{100}{1981}{37}; \np{216}{1983}{421}}
\refs{\lindeone,\ \langer,\ \affl,\ \lindethree}.
The nucleation temperature is the temperature where the rate per unit
volume of
critical
bubble nucleation equals the age of the universe multiplied by the
volume
inside the horizon. The critical bubble nucleation rate depends on
$F_c$, which
can be computed by finding a   spherically symmetric field
configuration
$\phi(r)$ which is a stationary point of the effective action and
has $\phi=0$
 for large $r$, and  $\phi$ near the minimum of the effective
potential for
small $r$. The free energy for such a configuration is approximately
(neglecting corrections to the derivative terms)
\eqn\crbub{
 S[\phi(r)]\approx\int r^2 dr\[{1\over 2}(\phi')^2  +V(\phi,T)
\] }
and a stationary point can be obtained by finding a nontrivial solution
of
\eqn\crdiff{{1\over r^2}{d\over d r}\(r^2\phi'\)-{\partial V\over
\partial
\phi}=0}
satisfying the boundary conditions
\eqn\bcond{\phi(\infty)=0, \qquad \phi(0)'=0\ .}
This must be done numerically, although an approximate analytic
formula is
given in ref.~\dlhll. In the minimal standard model with a Higgs mass
in the 30--80
GeV range the
result  is that the nucleation temperature is nearly  the
critical temperature~\refs{\lmt,\ \anha,\ \turok,\ \dlhll,\ \dhs}.
The  wall profile for the boundary between the two coexisting phases
at the critical temperature is the solution to the differential equation
\crdiff\ at   $T=T_c$ in the large $r$ limit, subject to the boundary
conditions   $\phi(\infty)=0$ and
$\phi(-\infty)=v$, where $v$ is the vev in the broken phase. This
solution should also give a good estimate for the profile of the wall of an
expanding bubble. (The exact wall shape of an expanding bubble in a thermal
medium is still unknown.)  If the effective potential at the critical
temperature is
$(\lambda/4)(\phi-v)^2\phi^2$, the analytic solution for a wall centered at
$r=0$ is
\eqn\wpro{\phi={v\over2}\[1+\tanh\(- r\;v \sqrt{\frac\lambda8}\;\)\]\
.}
The width of the wall $\delta_w\equiv(\sqrt{8/\lambda}\;)/v$  is
quite model dependent, which is unfortunate since, as we show in \S4,
this is an important parameter for calculating the baryon number produced.
In   the minimal standard model, if one takes the Higgs mass to have the
experimentally ruled out value 35 GeV so that $v/T\approx
1$, the wall width\foot{This estimate differs from ref.~\dlhll\
because of a factor of two difference in the definition of the wall width.}
is $\delta_w\sim 24/T$. In general   $\delta_w\propto v/\sqrt h$,  where
$h$ is the barrier   height between the   two minima of the effective
potential. In a model with   a strongly first order transition, such as a
version of the singlet   majoron model with the potential tuned to have a
nearly flat direction \kuy, or   a supersymmetric model with a gauge singlet
superfield \piet, the wall   can be thin, $\delta_w=\CO(1/T)$. These
calculations require  knowing the   effective potential in the region
between the two minima, and perturbation   theory breaks down at the
inflection points. In between the inflection points the perturbative
potential has an imaginary part, arising from the classical instability
of the homogeneous field configuration which we have assumed dominates   the
perturbative expansion \wewu. However the   lifetime
of such a homogeneous configuration is, in most cases, long enough that
it is meaningful to compute its free energy, which is approximately given
by  the real part of the effective potential \wewu.  Any smooth continuation
of  the effective potential from the regions where perturbation theory is
sensible will not deviate too far from the perturbative value, and for a
sufficiently strongly first order transition in a weakly coupled theory the
perturbative calculations should give a reasonable estimate for nucleation
rates  and the wall width $\delta_w$.

The calculation of the wall velocity $v_w$ is more complicated, since
it involves nonequilibrium interactions of the thermal plasma with the
wall. The higher pressure inside the bubble will tend to accelerate the
wall,   and the scattering of thermal particles from the wall will dissipate
energy   and slow the wall down \lindethree. A plausible assumption is that
the wall   will reach a terminal velocity where the different forces balance.
This assumption   was challenged by Turok \ref\fastwall{Turok N, Princeton
preprint   PUPT-91-1273 (1991)}, who pointed out that if the temperature is
spatially   constant and the particle distributions are always the local
thermal distributions for   a given value of $\phi$, then the force on the
wall is independent of   velocity and continuously accelerates the wall. This
is of course true, since the   wall can only slow down by dissipating
energy, which will cause a local   departure from equilibrium. Thus if the
wall is extremely thick when compared with   particle mean free paths, so
that the particle distributions approximately   equilibrate
with the changing value of $\phi$, $v_w$ will be ultrarelativistic.
For a  wall which is not infinitely thick, particle interactions with the
moving wall will   affect their distributions in a velocity dependent way.
For instance weak gauge   bosons and top quarks, which are massless in the
symmetric phase and  have large   masses in the interior of the bubble, will
tend to reflect from the wall. The   reflected particles absorb some of the
wall's momentum and the distribution of   reflected
particles is nonthermal. Several  estimates  of $v_w$, which include
the effects of particle interactions with the wall and rethermalization   in
various approximations, indicate  that $v_w\sim 0.1c$ for a wall of width
$\delta_w\simeq 1/T $  and is at most mildly relativistic for $\delta_w\simeq
(10-30)/T$\nref\khleb{Khlebnikov SYu, \physrev{46}{1992}{3223} }
\refs{\lmt,\ \dlhll,\ \khleb}. These calculations assume the minimal standard
model phase transition parameters with only $\delta_w$ allowed to vary; in
general $v_w$ is model dependent.

\goodbreak
\newsec{CALCULATING THE BARYON ASYMMETRY}

The master equation for weak scale baryogenesis is eq. \bvrate, which
may be rewritten as
\eqn\bmast{{dn_{_B}\over dt} =  -9 {\Gamma_a\over T} {\partial F
\over \partial
B}\equiv -9 {\Gamma_a \mu_{_B}\over T}}
where we have replaced $\Delta F = 3\partial F/\partial B$, the
derivative
computed at fixed $B-L$, remembering that $\Delta B = 3$ in an
anomalous event.
The rate $\Gamma_a$ is given in eqs. \aandm, \dbrate. We have given
above the reasons to suspect that two of the three necessary ingredients
for   baryogenesis may exist at the weak phase transition: baryon violation
and   departure from thermal equilibrium.  There remain the requirements of
$C$  and $CP$   violation, as well as some concrete mechanism that explains
how a nonzero   $\mu_{_B}$ can arise dynamically during the weak phase
transition.

\subsec{$CP$ violation}

We wish to explain the observed baryon to entropy ratio $n_{_B}/s
\simeq (0.6 - 1)\times 10^{-10} $ \kolbturn.   From the expression \dbrate\ for
the   anomalous rate $\Gamma_a$  we expect a factor of $\alpha_{w}^4 \approx
10^{-6}   $ in our final answer for $n_{_B}/s$, and   since $s=(2\pi^2
g_*/45)T^3$ with   $g_*\sim 10^2$ at the weak scale, one can only tolerate an
additional   suppression of
order $10^{-2}$ from the combined effects of $CP$ violating angles
$\delta_{CP}$, and various dynamical effects.   This guarantees that
$CP$ violating effects   must be large at weak scale energies, further
constraining possible models and giving reason to hope that new $CP$ violating
effects may be experimentally accessible beyond those seen in the kaon system.

It is natural to wonder whether the requisite $CP$ violation at the
weak scale could be of the Kobayashi-Maskawa (KM) form.  At high
temperature   one may expect that that a dimensionless measure of $CP$
violation $\delta_{CP}$ in the standard   model
could only depend on a reparametrization invariant combination of the
dimensionless Yukawa couplings and mixing angles. In \S1 we argued
that $\delta_{CP}\simeq 10^{-20}$,  a suppression factor  apparently
ruling out KM $CP$ violation as playing any role in weak
scale baryogenesis.  In the kaon system, the $CP$ violating parameter
$\epsilon$ is seventeen orders of magnitude larger than this estimate
because the $CP$ conserving contribution to $\bar{K^0}-K^0$ mixing is so
small.  It is conceivable that dimensionless numbers such   as $m_{pl}/T$ or
$\delta_w T$  could play a role in enhancing   $\delta_{CP}$ the
necessary eighteen orders of magnitude during baryogenesis.  Attempts
in this direction have been made in ref.~\shaposh, but the existence of these
enhancements remains unconvincing.  It is most likely that {\it any}
model of baryogenesis requires new sources of $CP$ violation beyond the
standard
model, and   that electroweak baryogenesis requires the new $CP$ violation to
be
at the weak   scale.

Incorporating additional $CP$ violation into the standard model
requires
 additional matter fields, and  there are a number of relatively
simple
extensions one can  consider.  Three examples that have been analyzed
in the
literature and are discussed below are the singlet Majoron model with
complex
phases in the neutrino mass matrix; the nonsupersymmetric two Higgs
doublet
model with a $CP$ violating relative phase between the  doublets; and
the
minimal supersymmetric standard model (MSSM) with a $CP$ violating
phase in the
gaugino masses.  In each of these models $\delta_{CP}$ can be
$10^{-2}$ or
larger, not being suppressed by small couplings as in the KM model,
and without
being in conflict with experimental limits on the electric dipole
moments (EDM)
of the electron or neutron.

\subsec{Timescales}
Having assembled the ingredients is not enough to make a universe: we
still
require a recipe.  The recipe involves the three distinct time scales
governing
the relevant reactions. These are the thermalization time $\tau_{_T}$
which
characterizes how fast particles in the cosmic plasma equilibrate;
the Higgs
time scale $\tau_{_H}$ which governs the departure from equilibrium,
given by
$\tau_{_H}\simeq \phi/\dot\phi$ as measured by a comoving observer
while the
expanding bubble wall passes through her; and the sphaleron time
$\tau_{sp}$
which governs the rate of baryon violation in the symmetric
phase\foot{ In
principle there is a fourth time scale, the age of
the universe $T/\dot T = H^{-1} $, but at the weak epoch it is many orders of
magnitude larger   than the
others and not relevant for microphysics.}.

Since  baryon violation  in the broken phase within the bubbles must
be very slow to prevent re-equilibration of the baryon number,
we will   assume here that baryon number is exactly conserved inside the
bubbles.
Outside   the bubbles in the symmetric phase, $\tau_{sp}$  at the transition
temperature $T$ is estimated as
\eqn\sprate{\tau_{sp}^{-1}\simeq\alpha_{w}^4 T = 1.3\times10^{-6} T\ .}
In contrast, the thermalization rate due to weak and strong
interactions, defined for the different particles as the inverse mean free path
$\ell_{_T}$,   is much faster, not having the $\alpha_{w}^4$ suppression. It is
estimated from strong or weak Coulomb scattering cross sections:
\dlhll
\eqn\thermrate{\taut^{-1}\equiv \ell_{_T}^{-1} \simeq
\cases{0.25\,T & q\cr 0.08\,T &
W,Z,$\ell$\cr} \ .}
These rates correspond to ``fast'' reactions; some standard processes
are much slower, of course, such as chemical equilibration between the first
and third families, or between the two chiralities of a light fermion.

The Higgs time scale is less easily determined, and is quite model
dependent, roughly given by
\eqn\hrate{\tauh^{-1}= \dot \phi / \phi \simeq {v_w\over \delta_w }
\simeq (0.01 - 1.0) T\ .}
We see that baryon violation is always out of equilibrium near the
wall, since
$\tau_{sp} \gg \tauh$. However, other particle interactions may or
may not be
able to equilibrate near the bubble wall, depending on the relative
size of
$\tauh$ and $\taut$. This gives rise to two distinct regimes:
\lfm
\item{1.} The adiabatic regime:  $\taut \ll \tauh$.  ``Fast''
interactions
maintain thermal equilibrium as the bubble wall passes by and the
value of the
Higgs field changes. This allows one to
to describe the plasma within the bubble wall in terms of
 equilibrium thermodynamics with
quasistatic chemical potentials for quantities that equilibrate
slowly compared
to $\tauh$.  In this
regime
baryogenesis occurs within the bubble wall itself.
\lfm
\item{2.} The nonadiabatic regime: $\taut \gg \tauh$.  The wall is
thin
compared to the mean free path and particles reflect off the oncoming
wall with
calculable $CP$ violating reflection coefficients. Baryogenesis
occurs in an
extended region preceding the phase boundary and is enhanced relative
to the
adiabatic scenario.
\lfm
Which regime one is in depends  on the bubble wall thickness and
velocity: for
thick or slow walls it is the adiabatic regime that is relevant; for
thin or
fast walls it is the nonadiabatic.

\subsec{The adiabatic ``thick wall'' regime: spontaneous
baryogenesis}

In the adiabatic limit it should be valid to treat the plasma in the
bubble
wall as being in quasi-static thermal equilibrium with a classical,
time
dependent field. The plasma will not be in chemical equilibrium,
however, as
some reactions, such as baryon violation, are slower than
$\tauh^{-1}$.  This
deviation from chemical equilibrium may be treated by introducing
chemical
potentials for the slowly varying quantities.

Baryogenesis arising from slowly varying classical  Higgs fields
during a phase
transition is called ``spontaneous baryogenesis''; it was first
discussed in
refs.~\nref\sponbar{Cohen AG, Kaplan DB, \pl{199}{1987}{251};
\np{308}{1988}{913}}\refs{\sponbar,\ \kolbturn} in the context of
second order
phase transitions, for which the assumption of adiabaticity is always
valid
(except in the extremely early universe).  To see how time dependent
Higgs
fields  can drive  baryon production consider a toy model with a
conserved
$U(1)_B$ symmetry carried by fermions $\psi\ (B=1)$, $\chi\ (B=-1)$, a
scalar
$\phi\ (B=2)$, and the Yukawa interaction $\bar\psi \phi\chi$. During
a phase
transition in which $\phi$ takes on the classical value
\eqn\phival{\phi(t) = \rho(t)e^{i\theta( t)}\ ,}
this leads to baryon production, even for $\dot \phi$ small compared
to the
fermion masses.  To see this, redefine the fermion fields by the
phase
$e^{i\theta( t)B/2}$; this removes the phase from the Yukawa
interaction, but
leads to a new interaction from the fermion kinetic term:
\eqn\cpot{\CL_{K.E.}\to \CL_{K.E.}+\half\dot \theta ( \bar \psi
\gamma^0\psi -
\bar \chi
\gamma^0\chi) = \CL_{K.E.}+\half\dot \theta n_{_B}\ .}
Thus $\dot \theta$ acts like a chemical potential for baryon number.
(We use
the word ``charge potential'' for $\dot \theta$ instead of chemical
potential,
since the effect is dynamical and not arising from a constraint.) The
interaction \cpot\ splits the energy levels of baryons versus
antibaryons so
that the free energy is minimized at nonzero baryon number. The
interaction
\cpot\ violates $C$  and $CP$ and will  spatially average to zero
except   in
theories with explicit violation of these symmetries.

If $B$ symmetry is violated in some other sector of the theory (\eg,
through
the electroweak anomaly), then the interaction \cpot\ will cause $B$
to try to
equilibrate to the  value $n_{_B} = \CO( \dot \theta T^2)$.  If the
$B$
violating interactions are rapid compared to $\ddot \theta/\dot
\theta$ then
this equilibrium value will be attained.  In the real case we wish to
study,
however,  the rate per unit volume of anomalous fluctuations
$\Gamma_a$ ( see
eq. \dbrate) is too slow to equilibrate due to the factor of
$\alpha_{w}^4$. In
this case one simply integrates $\dot n_{_B}$ in eq. \bmast\ over the
appropriate timescale.
It should be emphasized that it is not necessary for the charge
potential to
couple to the baryon current as in the toy example \cpot\ in order
for there to
be a nonzero value for $\dot n_{_B}$. As long as the charge $X$ is
not
orthogonal to $B$,  an interaction coupling of the form $\dot \theta
j^0_{_X}$
will  give rise  to a nonzero value for   $\mu_{_B}$  of the form
\eqn\muprop{  \mu_{_B} = \CN \dot\theta\ ,}
where $\CN$ is a calculable, model dependent constant.

After the bubble wall has passed by a given point in space, the
charge potential $\dot \theta$ returns to zero.  If baryon violation is
still  rapid compared to the cooling rate of the universe, the baryon
number   produced will re-equilibrate to $B=0$.  However, the anomalous
fluctuation rate   $\Gamma_a$
goes rapidly to zero as the Higgs vev turns on, so that the baryon
number drops out of chemical equilibrium and remains to the
present   epoch. By means of the inequality \transit\ we can make a crude
estimate of   this ``cut-off'' value of the Higgs vev:
\eqn\vevco{\vev{\phi}_{co} \simeq {7gT\over 2\pi} = 0.7\, T\ .}
To find the total baryon number produced during the phase transition,
eqs. \bmast, \muprop\  may be integrated across the bubble wall:
\eqn\btotex{n_{_B} = -{9\CN \over T}\int_{-\infty}^{t_{co}}\ {\rm d}t
\dot \theta \Gamma_a(\phi(t))\simeq -{9\CN \Gamma_a\over T}
\Delta \theta\   ,}
where we have estimated the integral by treating
$\Gamma_a$  as a   step function equal to its symmetric phase value for
$\vev{\phi}\le   \vev{\phi}_{co}$ and vanishing further inside the bubble.
The quantity $\Delta \theta$ is   then the change in $\theta$ from the
symmetric phase to the point in the   bubble wall where baryon violation
effectively shuts off;  it is a homogeneous   function of the $CP$ violating
parameter $\delta_{CP}$ of the theory in question. Furthermore, $\Delta
\theta$ is homogeneous in $\vev{\phi}_{co}$ and   could be very small if
$\vev{\phi}_{co}$ proves to be as small as suggested   in refs.
\refs{\dhss, \dhs}.

Given that the entropy density is $s=(2\pi^2 g_*/45)T^3$ and
$\Gamma_a = \kappa
( \alpha_w T)^4$, one finds for the final baryon to entropy ratio
\eqn\bsrat{{n_{_B}\over s} = -\kappa\({\CN\over 0.1}\)\({100\over
g_*}\)\({ \Delta\theta\over \pi}\)\times \(3\times 10^{-7}\)\qquad
{\rm (adiabatic)}\ ,}
to be compared to the observed value $n_{_B}/s \simeq 10^{-10}$. In
the above expression, the quantity  $\CN$  defined in eq. \muprop\ is
typically   $0.1 - 1.0$; $\kappa$ is the prefactor appearing in equation
\dbrate\  and   is thought to be $\CO(1)$; $(100/g_*)$ is also close to
unity at the weak phase transition.  The reason why this result is a formula
for the baryon   density of the entire universe, even though baryon
production is basically a   surface effect only occuring within the bubble
wall, is that the entire   universe is swept out by the bubbles during the
course of the phase transition   \usone.

We  now list the steps that must be followed
when analyzing weak scale baryogenesis models in the adiabatic limit, and
then proceed to give several examples.
\lfm
\item{(i)} Identify a model that has a first order phase transition
with large timescale $\tau_{_H}$ and classical Higgs field evolution during
the   phase transition such that there is a  $CP$ violating, time dependent
phase $\theta(t)$ in the Yukawa interactions.
\lfm
\item{(ii)}  Determine $\Delta \theta$ from the Higgs equation of
motion in order to use eq. \bsrat\ to compute the final baryon number.
Using   the generous estimate \vevco\ for $\vev{\phi}_{co}$,  $\Delta
\theta$ will   typically be $\CO(1)\times \delta_{CP}$.  Additional
suppression is possible if   baryon violation shuts  off closer to the outer
surface of the bubble and $\vev{\phi}_{co}$ proves to be much smaller than
assumed.
\lfm
\item{(iii)} Rotate the fermions in the theory to remove the time
dependent phase from the Yukawa interactions into a derivative coupling of
the   form $\dot \theta j^0_{_X}$. The current $X$ is ambiguous, since the
rotation is   not unique---one can always rotate so that $X\to (X+Q)$, where
$Q$ is a   classically conserved charge.  It is simplest to choose $X$ so that
the rotation   is anomaly-free, so one need not treat  $\theta(t) W\tilde W$
couplings. \lfm
\item{(iv)} Compute
the dynamically generated $\mu_{_B}$.  To do so one must introduce
chemical potentials for each conserved and appoximately conserved quantum
number. All the particle densities may be determined in terms of
these chemical potentials and the charge potential $\dot\theta$;  for
particle species $i$ the density $n_i$ is given by
\eqn\parden{ n_i = k_i\[X_i \dot\theta + \sum_a \mu_a
q^a_i\]{T^2\over 6}\ ,}
where $k_i$ is a statistical factor accounting for effective degrees
of freedom
  and differs for fermions and bosons, $X_i$ is the $X$ charge of
species $i$,
and the sum is over all charges out of chemical equilibrium
(including $B$).
The values of the chemical potentials $\mu_a$  are fixed by requiring
that the
primordial plasma not carry any  quantum numbers.  One can then
determine
$\mu_{_B}$ and the factor $\CN$ in eq. \muprop.
\lfm
\item{(v)} Use the
expression\ \bsrat\ to give the final baryon to entropy ratio.
 This formula is only expected to be valid deep in the
adiabatic
thick-wall regime, however.  If the equilibration time for particle
distributions is long compared with $\tau_{_H}$  there may be
additional
supression factors.
\lfm
We now present two examples of standard model extensions and show how
spontaneous baryogenesis proceeds in them.  The first is the two
Higgs doublet
model, and the second is the minimal supersymmetric standard model
(MSSM).

\lfm
{\it The two Higgs model.}
This model was first suggested by McLerran as a model for weak scale
baryogenesis in ref. \ref\mclax{McLerran L, \prl{62}{1989}{1075}}\
and analyzed
by Turok and Zadrozny in ref. \tz; the treatment here will follow our
subsequent work \usthree.  The model is characterized by the
Lagrangian
\eqn\lagth{ {\cal L}= \sum_{\hbox{fermions}} \bar\psi i \dsl\psi +
\sum_{i=1}^2
\vert D_\mu\phi_i\vert^2 -V(\phi_1,\phi_2) + \CL_{\rm yuk} +\CL_{\rm
gauge}}
where the scalar potential is given by  \ref\hhunt{Gunion JF, Haber
HE,  Kane
GL, Dawson S, {\bf The Higgs Hunter's Guide}, Addison-Wesley
Publishing
Company, (1990) }
\eqn\twoHiggspot{
\eqalign{V(\phi_1,\phi_2)&=\lambda_1 (\phi_1^\dagger\phi_1 -v_1^2)^2
+\lambda_2 (\phi_2^\dagger\phi_2 -v_2^2)^2 +
\lambda_3[(\phi_1^\dagger\phi_1-v_1^2)
 +(\phi_2^\dagger\phi_2-v_2^2)]^2 \cr
&+\lambda_4[(\phi_1^\dagger\phi_1)(\phi_2^\dagger\phi_2)-
(\phi_1^\dagger
\phi_2)(\phi_2^\dagger\phi_1)]+
\lambda_5[{\hbox{Re}}(\phi_1^\dagger\phi_2)-v_1v_2\cos\xi]^2\cr
&+ \lambda_6[{\hbox{Im}}(\phi_1^\dagger\phi_2)-v_1v_2\sin\xi]^2
\  .\cr}}
The Yukawa interactions in $\CL_{\rm yuk}$ couple the $\phi_1$ field
to up-type
quarks;  down-type quarks and charged leptons may be coupled
exclusively to
either $\phi_1$ or $\phi_2$---either choice suppresses flavor
changing neutral
currents and is protected by a discrete symmetry that is softly
broken. For
convenience, we will assume that all fermion masses arise due to
interactions
with $\phi_1$; the other option works equally well, since only the
top quark
Yukawa interaction proves to be relevant.  In either case we can
adjust the
phases of the Higgs doublets so that, in the fermion mass eigenstate
basis, the
top quark Yukawa interaction $y_t$ is real and $\phi_1$ develops a
real
expectation value at zero temperature. In this basis, aside from the
usual KM phase, there is an additional $CP$ violating phase $\xi$ in the
scalar potential, so that the zero temperature vev of the fields have a
relative phase $e^{i\xi}$ between them. During the weak phase transition the
neutral components of the two doublets will acquire time dependent classical
values with $\phi^0_1(t) = \rho(t)e^{i\theta(t)}$  (in unitary gauge where
the linear combination of the phases of $\phi^0_{1,2}$ eaten by the $Z$ has
been   rotated away).  The change in $\theta$ in going from the beginning of
the   phase transition to $T=0$ is \usthree
\eqn\dthet{
\Delta\theta=\CO\(\xi-\arctan\[\frac{\lambda_6}{\lambda_5}\tan\xi\]\)
.}
$\Delta\theta$ will be smaller in going from the symmetric side of
the domain wall to the point where the scalar vev  equals $\vev{\phi}_{co}$ and
baryon violation turns off, since $\Delta\theta$ scales like
$(\vev{\phi}_{co}/v)^2$,
arising from the change in relative importance between the $\phi^2$
and $\phi^4$ terms in the potential \twoHiggspot.

Next we remove  $\theta$ from the Yukawa interactions by performing a
time
dependent rotation of the fermion fields in the Lagrangian.  Since we
do not
wish to complicate our task and induce any coupling of $\theta$ to
gauge
fields, we  eliminate $\theta$ by means of an anomaly free fermion
rotation;
thus we rotate the fermions by an amount proportional to their
hypercharge. We
find that $\theta$ now has derivative couplings to twice the
fermionic part of
the hypercharge current (the ``$X$'' current in this model)
\eqn\model{\eqalign{
{\cal L}&= {\cal L}_{\hbox{K.E.}} -V(\phi_1,\phi_2)
 + y_t \bar q_L \tilde \rho_1 t_R + y_b \bar q_L \rho_1
b_R +\ldots +\hbox{h.c.}\cr
&+ 2\partial_\mu \theta \[\frac{1}{6}\bar q_L \gamma^\mu
q_L+\frac{2}{3}\bar U_R
\gamma^\mu U_R-\frac{1}{3}\bar D_R \gamma^\mu D_R-\frac{1}{2}\bar
\ell_L
\gamma^\mu \ell_L-  {\bar E}_R \gamma^\mu E_R \],
}}
where there is an implicit sum over the three families in the current
coupling
to $\partial_\mu \theta$, and $\rho_1$ is the radial component of
$\phi_1$. The
quantity  $\dot \theta$ which is nonzero during the phase transition
is then a
charge potential which splits fermion-antifermion energy levels
proportional to
their hypercharge inside the domain walls.

We now  must calculate the thermodynamic force
$\mu_{_B}=\CN\dot\theta$
responsible for the production of baryon number. To this end we
introduce
chemical potentials for each quantum number that equilibrates
slowly
compared to $\tau_{_H}$.  The relevant ones for this model are
$B-L$, $Q$, and
$B_3-\half (B_1+B_2)$ where $B_i$ is the baryon number of the
$i^{th}$ family.
(There are other conserved or slowly varying charges, such as
$B_1-B_2$, but
their chemical potentials turn out to be irrelevant.) From eq.
\parden\ we have
determined
\eqn\nvalth{\CN = {4\over 3}\({6+n\over 25+4n}\)}
where $n$ is the number of physical charged scalars which are light
compared to
the temperature $T$; for $n=0$, $\CN = 0.3$.  Comparing eq. \bsrat\
with
\dthet\ and \nvalth, we find that a value of $\xi\simeq
\CO(10^{-2}-10^{-3})$
for the $CP$ violating parameter in the theory yields an acceptable
value for
the final baryon to entropy ratio produced in the phase transition.

Our treatment is justified as long as fermionic hypercharge has time
to adjust
itself through the top quark mass or Yukawa interaction with thermal
Higgs
particles. This adjustment can occur via scattering (\eg, $t+ g \to
t+
\phi_1$), scalar bremstrahlung (\eg, $t+q\to t+ q +\phi_1$)  or
coherent top
mass oscillations within the domain wall. Comparing with photon
bremstrahlung \ref\bjd{Bjorken JD,
Drell SD, {\it Relativistic Quantum
Mechanics}, McGraw-Hill (1965)}\
we arrive at a crude estimate giving a fermionic hypercharge
equilibration rate
of $\tau_Y^{-1} \simeq \taut^{-1} \times (\lambda_t/\pi)^2 \simeq
0.03\,
T$, where $\taut^{-1}$ is the strong interaction thermalization
rate \thermrate.   Thus the above analysis is expected to be valid
for
$\tauh^{-1}$ at the low end of the range \hrate.  For  less
adiabatic
phase transitions there may be an additional suppression of our
result
proportional to $(\lambda_t/\pi)^2$ due to the inability of the
plasma to reach
complete chemical equilibrium in the time available.

A different treatment of the two Higgs model is found in refs. \tz,
\mstv,
\turokrev.   They find a contribution to the  effective action in an
expansion of
$m_t/T$ of the form
\eqn\mtsvop{ \Delta S = -i \zeta(3){7\over 2}\({m_t\over \pi v_1}\)^2
{1\over T^2} \int {\rm d}^4x \[ (\CD_i \phi_1^{\dagger} \sigma^a
\CD_0 \phi_1 +
{\rm h.c.}\] \epsilon^{ijk} F^a_{jk}}
which is argued to bias anomalous baryon violating transitions when
the Higgs
field $\phi_1$ takes on the classical value \phival; the computation
is shown
in detail in ref. \turokrev,  where an estimate of
$n_{_B}/s\simeq 10^{-10}\times (CP{\rm \ violating\ angles})$ is given.
This  computation
is subleading in the adiabatic limit, as it is proportional to
$(\lambda_t/\pi)^2$.  Furthermore the effect of this operator is more
strongly
affected by the value of $\vev{\phi}_{co}$ where baryon violation
cuts off, as
its effects scale like $\vev{\phi}_{co}^4 $  \turokrev, while the
factor of
$\Delta \theta$ in formula  \bsrat\ scales like
$\vev{\phi}_{co}^2$.

\goodbreak\lfm
{\it Minimal Supersymmetry.}   One of the most attractive extensions
of the
standard model is the minimal supersymmetric standard model (MSSM)
\hhunt. It
possesses two Higgs doublets, but does not follow the above
treatment, since
supersymmetry ensures that there be no $CP$ violation in the Higgs
potential
($\lambda_5 = \lambda_6$). Nonetheless, the MSSM has two possible
sources of
$CP$ violation, the phases conventionally called $\phi_A$ and
$\phi_B$, which
are absent in the minimal standard model~\ref\ssmcp{Ellis J,
Ferrara S, Nanopoulos DV, \pl{114}{1982}{231}; Buchmuller W, Wyler D,
\pl{121}{1983}{321}; Polchinski J, Wise MB,
\pl{125}{1983}{393};del Aguila F, Gavela M, Grifols J, Mendez A,
\pl{126}{1983}{71}; Nanopoulos DV, Srednicki M,
\pl{128}{1983}{61};Dugan M,
Grinstein B, Hall LJ, \np{255}{1985}{413}}; a combination of these
will contribute to baryogenesis.    The interactions in this model
are given in
terms of superfields by
\eqn\mssm{\eqalign{
&\left[\bar{U}\lambda_{U}QH
+\bar{D}\lambda_DQH'+\bar{E}\lambda_ELH'+\abs{\mu}e^{-i\phi_B}HH'
\right]_F \cr
&  +m_{3/2} \left[ \abs{A}e^{i\phi_A}(\bar{U}\xi_UQH
+\bar{D}\xi_DQH'+\bar{E}\xi_ELH') +\abs{\mu_B}HH' \right]_A \ ,\cr }}
along with mass terms for the gauginos. The $CP$ violating phases
$\phi_A$ and
$\phi_B$ occur only in interactions involving  the super-partners of
the
ordinary particles.

Suggestions for how spontaneous baryogenesis in the MSSM at the weak
phase
transition could proceed were first made in refs. \dhss\ and
\usthree; the
first  quantitative analysis is given in ref. \ussix.  There it was
found that
baryogenesis was negligibly affected by the phase $\phi_A$, but
proceeded due
to the effects of  the phase $\phi_B$ which appears in the mass
matrix of the
supersymmetric partners of the gauge bosons and the   Higgs scalars
(the ``{\it
inos}'').

The generalization of spontaneous baryogenesis from a single
fermion to
several fermions with a spacetime dependent mixing involves making a
space-time
dependent unitary change of basis on the fermions
\hbox{$\psi_{i}\rightarrow
U_{ij}(x_\mu)\psi_j$} in order to render  the fermion mass terms
everywhere
real, positive and diagonal;  the space-time dependence of $U$
requires that we
replace the kinetic energy terms in the Lagrangian by
\eqn\lkin{
{\cal L}_{\hbox{K.E.}} \rightarrow {\cal L}_{\hbox{K.E.}}
 + \bar\psi \gamma^\mu (U^\dagger  i\partial_\mu U) \psi
   \ .}
It is simplest to choose $U$   to be an anomaly free rotation, so
that there
are no $W\tilde W$ terms generated, which forces
$(U^\dagger  i\partial_\mu U)$   to couple to
ordinary quarks as well as   $inos$. In ref. \ussix\ it is shown that
the generalization of eq. \btotex\ is
\eqn\totbarnum{
n_{_B}={3\over T}\int^{\vev{\phi}_{co}}_0  d \phi  \Tr U^\dagger (\phi){
idU(\phi)\over d\phi}
t_{\rm wk}^2  \Gamma_a( \phi )  \ , }
where $\vev{\phi}_{co}$ is the value of the Higgs vev for which $\Gamma_a$
starts
falling exponentially (see eq. \vevco),
and $t_{\rm wk}^2$ is the $SU(2)$ Casimir
matrix.
Eq.~\totbarnum\ leads to an acceptable value for $n_{_B}/s$ for
certain regions
of parameter space in the MSSM.  However, between the restrictions
that the
phase transition be sufficiently first order and that $n_{_B}/s\simeq
10^{-10}$, a large region of parameter space is excluded, and the
theory is
experimentally testable as a baryogenesis candidate, a luxury
accorded to
few models.

\subsec{The nonadiabatic ``thin wall'' regime: the charge transport
mechanism}

Thin bubble walls occur naturally in models with scalar fields with
trilinear
couplings (\eg, a familiar extension of the MSSM), as well as in the
two Higgs
model or singlet Majoron with some tuning of parameters.  For bubble
walls
thinner than the strong interaction mean free path \thermrate,
$\delta_w<\ell_T\simeq 4/T$,  incoming fermions in the bubble wall
frame
interact with the wall like quantum mechanical particles scattering
from a
potential barrier.  Due to $CP$ violating interactions with the
scalar field,
the reflection coefficients for the fermions can be such that a net
charge is
reflected from the domain wall, which is transported into the region
preceding
the bubble.  This charge asymmetry (not necessarily in baryon number)
can bias
anomalous $SU(2)$ fluctuations in a large region in front of the
bubble wall
and produce a baryon asymmetry.  This ``charge transport mechanism''
was
introduced in the context of the singlet Majoron model, where the
fermions
scattering of the wall were massive neutrinos \usone;  here we will
summarize
ref. \usfour\ on the nonadiabatic limit of the two Higgs model
discussed above.
 Potentially the most compelling application of the nonadiabatic
regime would
be the MSSM extended by a singlet field \hhunt, but that model has
not been
analyzed to date.

The procedure that has been developed for analyzing the thin wall
scenario is
as follows:
\lfm
\item{(i)} Compute the wall velocity and profile, including the space
dependent
 $CP$ violating phase, from the finite temperature effective
potential.
\lfm
\item{(ii)} Calculate the reflection coefficients for fermions
striking the
wall with momentum $\vec k$ in the domain wall frame of reference by
integrating the Dirac (Majorana) equation of motion.  Because of $CP$
violation, the reflection probabilities for particles and
antiparticles can be
different.
\lfm
\item{(iii)} Convolute the reflection probabilities with the incoming
(boosted)
thermal flux of particles to determine the flux of fermions reflected
from the
wall.  Due to $CP$ violation, this flux will carry nonzero quantum
numbers $X$.
\lfm
\item{(iv)} The nonzero $X$ flux from the wall will lead to a nonzero
charge
density $n_{_X}$ in the region preceding the wall.  In general, a
nonzero
charge $X$ will cause the free energy to be minimized for nonzero
$B$, and one
finds
\eqn\dfdbx{\mu_{_B}={ \partial F\over \partial B}\Big\vert_{B=0} =
-\CC
{n_{_X}\over T^2}\ ,}
where $n_X$ is a nonzero $X$-charge density.  Given eq. \bmast, the
baryon
density in a region which is a distance $d(t)=(z-v_wt)$ in front of
the domain
wall equals
\eqn\nonadb{\eqalign{
n_{_B}(z,t) &= -{9\Gamma_a\over T} \int_{-\infty}^{z/v_w} {\rm d}t\
{\partial F\over \partial B}=\CC {9\Gamma_a\over
T}\int_{-\infty}^{z/v_w} {\rm
d}t\ n_{_X}(z-v_w t)\cr
&= \CC {9\Gamma_a\over T}{1\over v_w}\int_0^{\infty} {\rm d}z\
n_{_X}(z) \ .}}
Thus one needs to calculate the total amount of charge in the region
in front
of the wall in order to compute the net baryon density produced. (As
in the
adiabatic scenario, baryons that are produced pass into the broken
phase and
become stable; a major difference though is that the baryon asymmetry
is
produced in front of the bubble in the symmetric phase, rather than
within the
bubble wall.)

\lfm
The applications of this procedure to the two Higgs model is detailed
in ref.
\usfour.  Rather than  compute the bubble wall parameters from the
effective
potential, we use a representative wall profile
\eqn\wallprof{  \phi_1(z)= v_1 \[ {1+\tanh(z/\delta_w)\over 2}\]
e^{-i\Delta\theta [\tanh(z/\delta_w)-1]/2}\ ,}
where the top mass $m_t(T) = \lambda_t v_1$ and the wall width
$\delta_w$ are
taken as free parameters to vary, and  $\Delta\theta=-\pi$
is assumed;
the resultant baryon asymmetry should scale
linearly for
small  $\Delta \theta$. The only fermion that has appreciable
reflection
probability from the domain wall is the top quark, and $CP$ violation
causes
an asymmetry in the reflection rate between right-handed and
left-handed top
quarks, leading to a flux carrying net axial charge shown in
\fig\asym{The
difference in reflection probabilities for left and right handed top
quarks, as
a function of the longitudinal momentum in the wall frame, for three
different
wall widths. The wall shape is given in \wallprof, with
$\Delta\theta=-\pi$,
corresponding to maximal $CP$ violation.}.  The asymmetry is
maximized for
particles whose longitudinal momentum in the wall frame is of order
the top mass in the broken phase at that temperature (all particles
get
reflected or transmitted, if their longitudinal momenta are lower
 or much
higher than $m_t$, respectively).  From these $\vec k$ dependent reflection
probabilities
the momentum distribution of the reflected flux is computed.
The
transport of the axial top quark number in front of the wall is
calculated by
Monte Carlo techniques, allowing each reflected top quark to undergo
Coulomb
scattering via gluon exchange (with an infrared cutoff in the form
of a
plasmon mass) until the wall catches up.  The
proportionality factor
$\CC$ in eq. \dfdbx\ is found to equal $4/(1+2n)$ in the standard
model
augmented to have $n$ Higgs doublets, and from eq. \nonadb\ we
compute the baryon asymmetry.

Axial top number is not a good quantum number, as it is violated by
Higgs
scattering, and so in ref. \usfour\ fermionic hypercharge was
identified as
the transported charge.  In \fig\densf{A plot of the ratio of
hypercharge to
entropy densities preceding the bubble wall, as seen in the thermal
frame of
reference. The horizontal axis is the distance ahead of the wall, in
units of
the inverse temperature.  These computations were performed assuming
$m_t=2 T$
and $\delta_w= 1/T$, for the two sample wall velocities
$v_w=1/\sqrt{3}$
\hbox{($\gamma=1.225$)} and $v_w=.98$ \hbox{($\gamma=5.0$)}. It is
the area
under these curves that enters eq. \nonadb\ for $n_{_B}$.} we display
the
results of the Monte Carlo simulation for several parameters, showing
a
substantial transport of charge in front of the domain wall.  The
penetration
distance of the charge pushed in front of the wall depends on the
velocity of
the wall, and for $v_w = 1/\sqrt{3}$---the sound velocity---we find a
distance
of $\sim 40/T$; for slower walls, the distance is greater.  There
are several
factors contributing to this large distance:  (i) The incident
particles
selected by the reflection coefficients have longitudinal momentum
$p_L\sim
m_t=2T$, and so have higher than average energy to start with; (ii)
Upon
reflection, the particle energies are enhanced by $\gamma (
1+\beta)\simeq 2$,
for $v_w = 1/\sqrt{3}$; (iii)  Particles typically
experience
numerous collisions, but primarily in the forward direction. Our
numerical
results are consistent
 with Bjorken's energy loss formula \ref\eloss{Bjorken
JD,
FERMILAB-Pub-82-59-THY (1982); Braaten E, Thoma M,
\physrev{44}{1991}{R2625}}\
for light quarks in a quark-gluon plasma
\eqn\bjform{{{\rm d}E\over {\rm d} x} = - {8\pi\over 3} \alpha_s^2
(1+N_f/6) T^2 \ln\sqrt{4ET/M^2}\ ,} where $N_f=6$ is the number of
quark flavors in the plasma and $M$ is an infrared cuttoff ($\simeq$ the
plasmon mass): this formula  yields a stopping distance of $\ell\simeq
(9/8\pi \alpha_s^2) T^{-1} = 35 T^{-1}$ for a reflected top quark with
initial energy $E\simeq 6T$, twice the average thermal energy. (iv) Finally,
for   very slow domain walls, diffusion of charge in front of the wall plays
a large   role and increases the penetration distance substantially.

Khlebikhov has shown by means of the Vlasov equations that the above
scenario cannot proceed exactly
as described above, since hypercharge is gauged and
any hypercharge density produced will be screened by the plasma
\ref\khleb{Khlebnikov SYu, UC Los Angeles preprint   UCLA-92-TEP-14
(1992) }.
In ref. \usfive\ this error was corrected.  The correct charge to
consider as the ``$X$'' charge in eq. \nonadb\ is not fermionic hypercharge
$Y_f$, but
rather the linear combination $B' = B-xY_f$ where $B$ is baryon
number and $x$ is chosen so that $B'$ is not screened by gauge  forces; we
find $x=2/(10+n)$, where $n$ is the number of scalar   doublets. The
reflected flux of top quarks carries $B'$ charge, and upon redoing
the analysis of ref. \usfour, the final results for the produced baryon
asymmetry are found to be unchanged.  These are shown in
\fig\nbcont{Contour   plot of
$n_{_B}/s$ as a function of the top mass and $\delta_w$, in units of the
critical temperature, assuming maximal   $CP$ violation ($\Delta\theta=-\pi$
in eq. \wallprof).   We have  assumed   the number of light scalars to be
$n=2$, the wall velocity to be $v_w =   1/\sqrt{3}$ and have taken
$\kappa=1$ in eq. \dbrate.}.  The contour plot shows that a   very large
baryon asymmetry can result for maximal $CP$ violation, $v_w = 1/\sqrt{3}$
and $\delta_w \times m_t =\CO( 1)$, where $m_t$ is the   value of the top
mass in the broken phase at the phase transition. For a wall   velocity $v_w
= 0.1$ one finds an order of magnitude enhancement in $n_{_B}/s$   \usfour.

We conclude that for thin bubble walls the charge transport mechanism
works very well for producing the baryon asymmetry, and in fact
suffers from fewer theoretical uncertainties than the adiabatic scenarios
where   baryogenesis actually takes place within the bubble wall.

\subsec{Conclusions about mechanisms}
As is evident from the above discussion, the mechanism for weak scale
baryogenesis depends on the width and velocity of the bubble wall.
Allowing for
certain processes to remain in thermal and chemical equilibrium
during the phase transition, we have been able to show how weak scale
baryogenesis could proceed in both the adiabatic and nonadiabatic
regimes,
producing a phenomenologically acceptable value for $n_{_B}/s$.  It
is tempting
to ``interpolate'' and assume that baryogenesis would also work for
the
interesting case of phase transitions that lie between these two
regimes,
where scattering lengths are comparable to interaction lengths with
the domain
wall;  however, the tools of equilibrium thermodynamics do not
suffice for
such a regime which would have to be analyzed by means of the
Boltzman
equation. Such a treatment could  afford  detailed knowledge
about the crucial departure from equilibrium in all regimes.

In addition to the models discussed above there are a number of
others in the literature, which we include for completeness.  An
early model of
McLerran \mclax\ attempted to use cosmological axions to provide the
required
$CP$ violation in standard model baryogenesis, but the resultant
baryon
asymmetry proved to be too small. Other weak scale baryogenesis
investigations
include the effect of anomalous QCD events at the weak scale
\ref\mzone{
Mohapatra RN, Zhang X, \physrev{45}{1992}{2699}};  left-right
models~\nref\mztwo{Mohapatra RN, Zhang X, University of Maryland
preprint
UMDHEP-92-230 (1992)}\nref\fhmot{Fr\`ere J-M, Houart L, Moreno JM,
Orloff J,
Tytgat M, CERN preprint TH 6747/92 (1992)}\refs{\mztwo,\ \fhmot};
transient
scalar-top condensation in supersymmetric models \ref\sher{Sher M,
William \&
Mary College preprint WM-92-107 (1992)}; lepton number
production from domain walls colliding at the end of the
transition~\ref\mari{ Masiero A, Riotto A,
\pl{289}{1992}{73}}; and
baryogenesis from   electroweak
strings without a first order phase transition
\ref\branden{Brandenberger RH, Davis AC, Brown University preprints
BROWN-HET-862, BROWN-HET-865 (1992)}.

\newsec{CONSTRAINTS AND EXPERIMENTAL SIGNATURES}

One of the most appealing features of weak scale baryogenesis is the
implication that new physics lies in wait for us at the weak scale.
Since weak
scale baryogenesis is feasible in most extensions of the standard
model
experimental guidance is needed.

Possible and expected experimental consequences of weak scale
baryogenesis are:

\lfm
{\it 1. New  particles.}  The requirement that the weak phase
transition be sufficiently first order, combined with Higgs mass limits from
LEP,   imply that the weak symmetry breaking sector be extended beyond a
single Higgs   doublet. Furthermore, the requirement that there be new $CP$
violation also   requires  that new particles be added to the standard
model.  Both constraints   suggest that there ought to be more than one
scalar particle for the SSC to   find.

\lfm
{\it 2. New $CP$ violation.} The new sources of $CP$ violation may be
large enough to see in future experiments measuring the electric dipole
moment of the electron and neutron \ref\dipole{ Barr SM, Marciano WJ, in {\it
CP   Violation},
ed. C. Jarlskog (World Scientific, Singapore, 1989) Bernreuther W,
Suzuki M,
\rmp{63}{1991}{313}}, atomic physics
\ref\atomic{Barr
SM, \prl{68}{1992}{1822}; \physrev{45}{1992}{4148}}, or top quark
production \ref\topcp{ Atwood D,  Soni A, \physrev{45}{1992}{2405},
Schmidt CR, Peskin ME, \prl{69}{1992}{410};
Schmidt CR,
\pl{293}{1992}{111}}.

\lfm
{\it 3. Constraints on supersymmetry.}  Spontaneous baryogenesis at
the weak scale in the MSSM  puts significant constraints on the
experimentally
allowed region of parameter space.  For example  the minimal
supersymmetric standard model can avoid baryon number washout
provided that the
lightest Higgs scalar is lighter than about 50 GeV, $\tan\beta$,
the ratio of
the two Higgs vevs, is smaller than about 1.7, the top quark is
heavier than
about 150 GeV, and the squarks are lighter than about 150 GeV \myint.
(All of
these bounds are only approximate because they are computed in   one
loop
perturbation theory, and can be weakened by allowing the top squark
to be
significantly lighter than the other squarks.) Creating sufficient
baryon
asymmetry requires    the electric dipole moment of   the neutron to
be greater
than $10^{-27}$ e-cm \ussix.
\lfm
{\it 4. Flavor changing neutral currents.}  Typical extensions of the
standard model that are appropriate for baryogenesis  (\eg the two
Higgs
model) could have significant flavor changing neutral currents large
enough to
see in the heavy flavor sector.  Interesting  places to look are
$B^0\to
\mu^+\mu^-$ \ref\msbmm{Savage MJ, \pl{266}{1991}{135}},
$t\to c \gamma$, and   $t\to c
Z$
\ref\mlms{Luke M, Savage MJ, UC San Diego preprint UCSD-PTH-93-01}.
\lfm
{\it 5. Dark matter searches.}
Anomalous weak processes could explain the relative abundance of dark
matter to baryon number in the universe today by coproduction of massive
stable particles along with baryons through the anomaly \ref\dark{Barr SM,
Chivukula SR, Farhi E, \pl{241}{1990}{387}; Barr SM, \physrev{44}{1991}{3062};
Kaplan DB, \prl{68}{1992}{741}}.  These particles would have to carry weak
interactions, and are tightly constrained by present dark matter searches
\ref\cald{ Ahlen SP
\etal, \pl{195}{1987}{603}; Caldwell DO \etal, \prl{61}{1988}{510}}.

\goodbreak
\newsec{FUTURE CALCULATIONS}
We have seen that the baryon number of the universe may have been
produced at the weak phase transition via experimentally accessible weak scale
physics. A precise calculation of the baryon abundance produced in a given
model would allow us to pin down the parameters in that model, much as the
abundances of the light elements and of dark matter are used to constrain
models.   So far, the baryogenesis calculations have been qualitative, and
probably only   accurate to
within a couple of orders of magnitude. A more quantitative
calculation  will require further analytic and numerical work on the
following questions:
\lfm
\item{1.} A lattice study of the order and parameters (critical temperature,
value of   the order parameter at the critical temperature) of the electroweak
phase   transition.   It is is of particular
interest to see how the parameters depend on the Higgs mass.
\lfm \item{2.} A computation of the baryon violation rate in both
phases and within the bubble wall itself.
\lfm \item{3.} A quantitative calculation of the nonequilibrium
transport of quantum numbers is needed both for all but the most adiabatic
cases,   for instance by numerically solving the Boltzmann equations.

\bigskip\bigskip \centerline{{\bf Acknowledgements}} The
work of A.N. and D.K.  was supported in part by the Department of
Energy under
contract \#DE-FGO3-90ER40546 and fellowships from the Alfred P. Sloan
Foundation.  D.K. is also supported in part by the NSF under contract
PHY-9057135. The work of A.C. was supported in part by the Department
of Energy
under contract \#DE-FG02-91ER40676 and by the Texas
National
Research Laboratory grant \#RGFY92B6.
\vfill\eject
\listrefs
\vfill\eject
\listfigs
\end
\bye